\documentclass[a4paper, 12pt]{article}  
\usepackage{pdfpages}
\usepackage{lmodern}
\usepackage{graphicx} 
\usepackage{wrapfig} 
\usepackage{textcomp} 
\usepackage[utf8]{inputenc}
\DeclareSymbolFont{letters}{OML}{ztmcm}{m}{it}
\usepackage{color}
\usepackage{bm}
\usepackage{multirow}
\usepackage{listings}
\usepackage{adjustbox}
\usepackage[boxed]{algorithm2e}
\usepackage{subcaption}
\usepackage{multirow} 
\usepackage[OMLmathrm,OMLmathbf]{isomath}
\usepackage{booktabs} 
\usepackage[hang,flushmargin]{footmisc}
\usepackage{pdflscape}
\usepackage{natbib}

\usepackage{datetime}
\usepackage[T1]{fontenc} 
\usepackage{geometry}
\usepackage{amsmath}
\usepackage{etoolbox} 
\usepackage{amsthm}
\usepackage{amsfonts}
\usepackage{amssymb}
\usepackage[section]{placeins}
\usepackage{diagbox} 
\usepackage{hyperref}
\usepackage{graphicx} 
\usepackage{fixmath} 
\usepackage[scaled=.90]{helvet}
\usepackage{wrapfig}
\usepackage{tikz}
\makeatletter
\renewcommand\@biblabel[1]{\textbf{#1.}} 
\usepackage{enumitem}

\usepackage{pifont}
\setlist{nolistsep,leftmargin=*}
\renewcommand{\maketitle}{ 
	\begin{center}
		{\LARGE\@title} 
		
		\vspace{0pt} 
		
		{\large\@author} 
		\\\@date 
		
		\vspace{40pt} 
	\end{center}
}
\begin{document}

\begin{center}
	\textbf{Statistical modelling of COVID-19 data: Putting Generalised Additive Models to work} \\
	Cornelius Fritz, Giacomo De Nicola, Martje Rave, Maximilian Weigert,  \\ Yeganeh Khazaei, Ursula Berger,     
	Helmut Küchenhoff, G{\"o}ran Kauermann \\
Ludwig-Maximilians-Universität München\hspace{.2cm}
\end{center}

\begin{abstract}
    Over the course of the COVID-19 pandemic, Generalised Additive Models (GAMs) have been successfully employed on numerous occasions to obtain vital data-driven insights. In this paper we further substantiate the success story of GAMs, demonstrating their flexibility by focusing on three relevant pandemic-related issues. First, we examine the interdepency among infections in different age groups, concentrating on school children. In this context, we derive the setting under which parameter estimates are independent of the (unknown) case-detection ratio, which plays an important role in COVID-19 surveillance data. Second, we model the incidence of hospitalisations, for which data is only available with a temporal delay. We illustrate how correcting for this reporting delay through a nowcasting procedure can be naturally incorporated into the GAM framework as an offset term. Third, we propose a multinomial model for the weekly occupancy of intensive care units (ICU), where we distinguish between the number of COVID-19 patients, other patients and vacant beds. With these three examples, we aim to showcase the practical and ``off-the-shelf'' applicability of GAMs to gain new insights from real-world data. 
\end{abstract}

\section{Introduction}

From the early stages of the COVID-19 crisis, it became clear that looking at the raw data would only provide an incomplete picture of the situation, and that the application of principled statistical knowledge would be necessary to understand the manifold facets of the disease and its implications \citep{Panovska-Griffiths2020, Pearce2020}. 
Statistical modelling has played an important role in providing decision-makers with robust, data-driven insights in this context. In this paper, we specifically highlight the versatility and practicality of Generalized Additive Models (GAMs). GAMs constitute a well-known model class, dating back to  \cite{Hastie-Tibshirani:1987}, who extended classical Generalized Linear Models \citep{Nelder1972} to include non-parametric smooth components. This framework allows the practitioner to model arbitrary target variables that follow a distribution from the exponential family to depend on covariates in a flexible manner. 
Due to the duality between spline smoothing and normal random effects, mixed models with Gaussian random effects are also encompassed in this model class \citep{Kimeldorf1970}. One can justifiably claim that the model class is one of the main work-horses in statistical modelling (see \citealp{WOOD:2017} and \citealp{Wood2020} for a comprehensive overview of the most recent advances). Numerous authors have already used this model class for COVID-19-related data analyses. As research on topics related to COVID-19 is still developing rapidly, a complete survey of applications is impossible; hence, we here only highlight selected applications, sorted according to the topic they investigate. Numerous applications analyse the possibly non-linear and delayed effect of meteorological factors (including, e.g., temperature, humidity, and rainfall) on COVID-19 cases and deaths (see \citealp{GOSWAMI2020801,PRATA2020138862,https://doi.org/10.1111/tbed.13631,XIE2020138201}). While the results for cold temperatures are consistent across publications in that the risk of dying or being infected increases, the findings for high temperatures diverge between studies from no effects \citep{XIE2020138201} to U-shaped effects \citep{MA2020138226}. Logistic regression with a smooth temporal effect, on the other hand, was used to identify adequate risk factors for severe COVID-19 cases in a matched case-control study in Scotland \citep{McKeigue2020}. In the field of demographic research, \citet{Basellini2021} investigate regional differences in mortality during the first infection wave in Italy through a Poisson GAM with Gaussian random effects that account for regional heterogeneities. With fine-grained district-level data, \cite{FritzKauermann:21} present an analysis confirming that mobility and social connectivity affect the spread of COVID-19 in Germany. \cite{Wood2021} shows that UK data strongly suggest that the decline in infections began before the first full lockdown, implying that the measures preceding the lockdown may have been sufficient to bring the epidemic under control. This list of applications illustrates how GAMs have been successfully employed to obtain data-driven insights into the societal and healthcare-related implications of the crisis. 


We contribute to this success story by focusing on three applications to demonstrate the ``off-the-shelf'' usability of GAMs. First, we investigate how infections of children influence the infection dynamics in other age groups. In this context, we detail in which setting the unknown case-detection ratio does not affect the (multiplicative) parameter estimates of interest. Second, we show how correcting for a reporting delay through a nowcasting procedure akin to that proposed by \citet{Lawless1994} can be naturally incorporated in a GAM as an offset term. Here, the application case focuses on the reporting delay of hospitalisations. Third, we propose a prediction model for the occupancy of Intensive Care Units (ICU) in hospitals  with COVID-19 and non-COVID-19 patients. We thereby provide authorities with interpretable, reliable and robust tools to better manage healthcare resources. 

The remainder of the paper is organised as follows: Section \ref{sec:data} shortly describes the available data on infections, hospitalisations and ICU capacities that we use in the subsequent analyses, which are presented in Sections \ref{sec:modeeling_infection_dynamics_across_age_groups}, \ref{sec:reporting_delay_hospitalization} and \ref{sec:modeling_icu}, respectively. We conclude the paper in Section \ref{sec:discussion}.

\section{Data}
\label{sec:data}
For our analyses, we use data from official sources, which we describe below. Note that our applications are limited to Germany although all of our analyses could be extended to other countries given data availability. We pursue all subsequent analyses on the spatial level of German federal districts, which we henceforth refer to as ``districts''. This spatial unit corresponds to NUTS 3, the third and most fine-grained category of the NUTS European standard (Nomenclature of Territorial Units for Statistics). We refer to Annex A for a graphical depiction of the spatial resolution of the data. 


 \paragraph*{Infections and hospitalisations}
 For investigating infection dynamics across different age groups, we use data provided by the Bavarian Health and Food Safety Authority (Landesamt für Gesundheit und Lebensmittelsicherheit, \href{https://www.lgl.bayern.de/}{LGL}). This statewide register includes, the registration date for all COVID-19 infections reported in Bavaria, as well as information on the patent's age and gender. Infection data for Germany is also published daily by the RKI \citep{Rki2021}, the German federal government agency and scientific institute responsible for health reporting and disease control. Due to privacy protection, the RKI groups patients in broad age categories, which inhibits the analysis of the group of school children. As this is necessary for our first application in Section \ref{sec:infection_dynamics_application}, we restrict the analysis to Bavarian data and use LGL data where not stated otherwise. 
 
  In addition, the LGL dataset includes information on the hospitalisation status of each patient, which is not included in the RKI data, i.e.\ whether or not a case has been hospitalised and the date of hospitalisation, if this had occurred. We determine the date on which a hospitalised case is reported to the health authorities by 
  matching the cases across the downloads available on different dates. This is necessary in order to derive the reporting delay for each hospitalisation, which is of interest in Section \ref{sec:reporting_delay_hospitalization}.
  
 

\paragraph*{Intensive care unit occupancy}
Data on the daily occupancy of ICU beds in Germany, on the other hand, is made publicly available by the German Interdisciplinary Association for ICU Medicine and Emergency Medicine (Deutsche interdisziplinäre Vereinigung für Intensiv- und Notfallmedizin, \citep{Divi2021}). Using this dataset we obtain information on the number of high and low care ICU-beds occupied by patients infected with COVID-19 and patients not infected with COVID-19. As a third category, there are also the vacant beds. In contrast to the infection data, no information is available on the age or gender composition of the occupied beds. 

\paragraph*{Population Data}
In conjunction with the data sources described above, we use demographic data on the German population at the administrative district level, provided by the German Federal Statistical Office (DESTATIS). Since the raw numbers on infections and hospitalisations are strongly influenced by the number of people living in a particular district, we use this population data to transform the absolute infection and hospitalisations to incidence rates. In general, we use the term incidence rates to refer to infection incidence rates, and hospitalisation incidence rates when writing about hospitalisations. While we effectively model the incidence rate in Section \ref{sec:modeeling_infection_dynamics_across_age_groups} and the hospitalisation incidence rate in Section \ref{sec:reporting_delay_hospitalization}, we incorporate the incidence rate per 100.000 inhabitants as a regressor in Section \ref{sec:modeling_icu}.   

\section{Analysing associations between infections from different age groups}
\label{sec:modeeling_infection_dynamics_across_age_groups}

A central focus during the COVID-19 pandemic is to identify the main transmission patterns of the infection dynamics and their driving factors. In this context, the role of children in schools for the general incidence poses an important question with many socio-economic and psychological implications to it  (see \citealp{Andrew2020,Luijten2021}). Since findings from previous influenza epidemics have tended to identify the younger population, children aged between 5 and 17, as the key ``\textsl{drivers}'' of the disease \citep{Worby2015}, the German government ordered school closures throughout the course of the pandemic between spring 2020 and 2021 to contain the pandemic. However, whether these measures were necessary or effective in the case of COVID-19 is still subject to current research \citep[e.g.][]{Perra:21}. In particular, several studies investigated the global effect of infections among school children, but a general conclusion could not be drawn (see \citealp{Flasche2021,Hippich2021,Hoch2021,ImKampe2020}). In general, we would like to remark that in many studies the main goal was to arrive at conclusions about the susceptibility, severity, and transmissibility of COVID-19 for children\citep{Gaythorpe2021}. On the other hand, we are here primarily interested in quantifying how the incidences of children are associated with the incidences in other age groups. Therefore, we want to assess whether children are key ``\textsl{drivers}'' of the pandemic. Our analysis is based on aggregated data on the macro level, as opposed to the data on the individual level, which is needed to answer hypotheses , e.g., about the susceptibility of a particular child.       

\subsection{Autoregressive model for incidences}
\label{sec:Autoregressive}

\begin{figure} [t]
    \centering
    \includegraphics[width=0.6\linewidth, page =1]{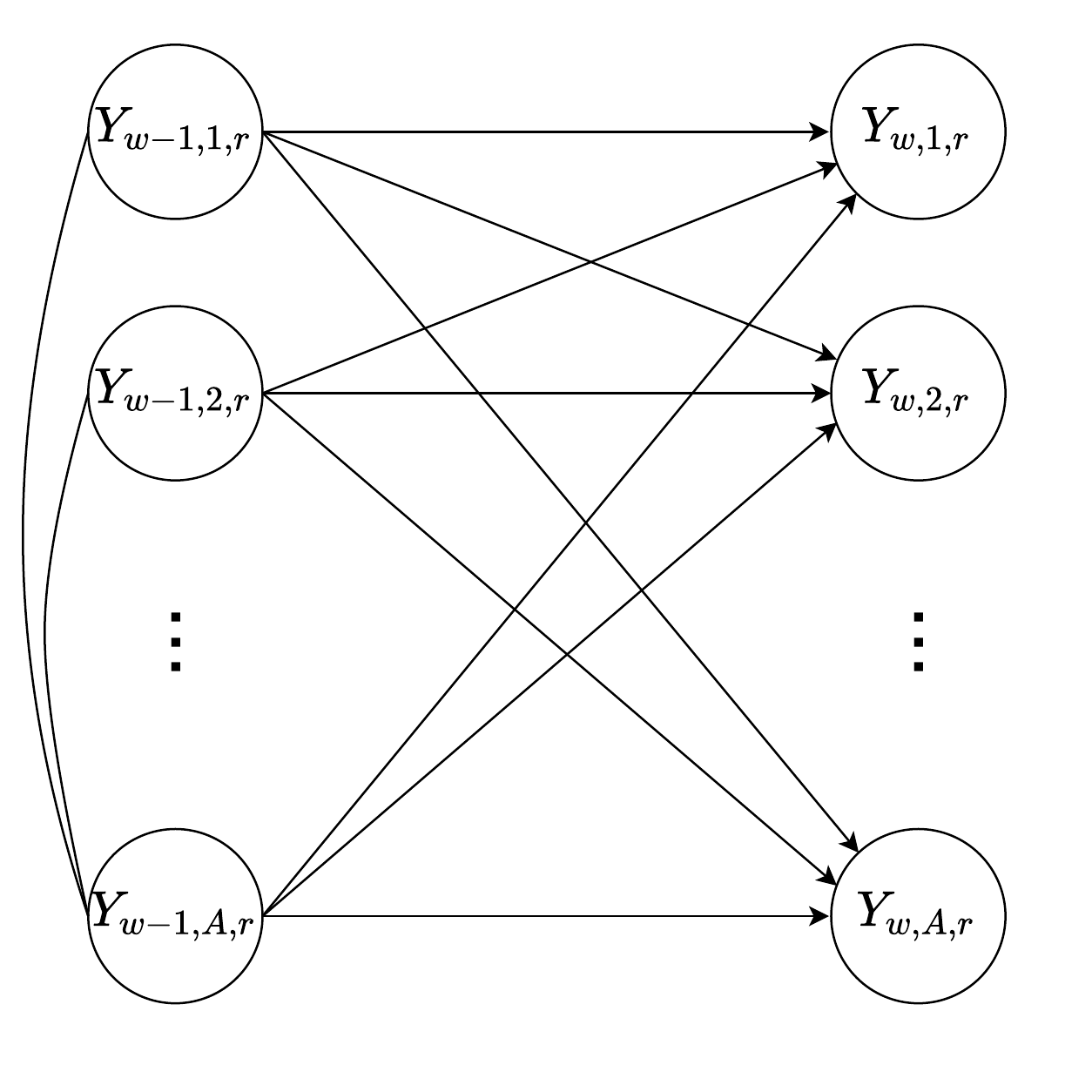}
    \caption{Assumed temporal dependence structure visualised as directed acyclic graph (DAG).}
    \label{Fig:DAG}
\end{figure}

To tackle this problem from a statistical point of view, we propose to analyse the infection data using a time-series approach \citep{Fokianos2004}. Let therefore $Y_{w,r,a}$ denote the number of reported infections in week $w$ in district $r$ and age group $a$. For simplicity, we assume independent developments among the districts and let $Y_{w, r, a}$  depend on the incidences in all age groups from  the previous week $w-1$. Put differently, we include $ Y_{w-1,r} = (Y_{w-1, r , 1}, … , Y_{w-1, r, A})$ as covariates, where $1, ..., A$ indexes all $A$ considered age groups. Among the components of $Y_{w, r }$ we then postulate independence conditional on $Y_{w-1, r}$. For illustration, Figure \ref{Fig:DAG} depicts the assumed dependence structure. 
As for the distributional assumption, we make use of a negative binomial distribution with mean structure
\begin{equation} \label{eq:mean}
\mathbb{E}(Y_{w, r, a} \vert Y_{w-1, r}) = \exp \{ \eta_{w,r,a} + o _{r, a} \} 
\end{equation}
where $o _{ r, a}$ serves as offset and $\eta$ gives the linear predictor. To be specific, we set $o_{r, a} = \log(x_{\text{pop},r,a})$, where $x_{\text{pop},r,a}$ is the time-constant population size in district $r$ and age group $a$. Note that we implicitly model the incidences by incorporating this offset term, since the incidences ${I}_{w, r, a}$ relate to the counts through $Y_{w, r, a} ={I}_{w, r, a}x_{\text{pop},r,a}$. The linear predictor is now defined as 
\begin{equation}
 \eta_{w,r,a} =  \theta_{w} + \sum^A_{k=1} \log(Y_{w-1, r, k} + \delta) \theta_{a,k},
  \label{eq:lp_ts}
\end{equation}
where $\theta_{w}$ serves as week-specific intercept and $\delta$ is a small constant, which is included for numerical stability to cope with zero infections.  We set $\delta$  to 1 in the calculation but omit the term subsequently for a less cluttered notation.

\subsection{Robustness under time-varying case-detection ratio}
\label{sec:robustness}

Model (\ref{eq:mean}) has the important methodological advantage of being able to cope with an unknown case-detection ratio, which is inevitable if there are under-reported cases. This is a key problem in COVID-19 surveillance as not all infections are reported \citep{Li2020b}; hence the case-detection ratio (CDR) is typically less than one. Various approaches have been pursued to quantify the number of unreported cases, e.g., by estimating the proportion of current infections which are not detected by PCR tests \citep{Schneble-etal:21}. For demonstration, assume that $\tilde{Y}_{w,r, a}$ are the detected infections in week $w$ in district $r$ for age group $a$, while $Y_{w,r,a}$ are still the true infections. Apparently $\tilde{Y}_{w,r, a} \leq Y_{w,r,a} $ holds if we assume   under-reporting. We assume multiplicative under-reporting and  denote with  $ 0< R_{w,r, a} \leq 1$ the multiplicative CDR  in district $r$ in age group $a$ and set with $ R_{w,d} = ( R_{w,r, 1}, ...,  R_{w,r, A})$ the joint CDRs for all $A$ available age groups. In this setting, we observe 
\begin{equation} \label{eq:dark}
\tilde{Y}_{w,r, a} = R_{w,r, a} Y_{w,r, a}
\end{equation}
infections in the corresponding week $w$, district $r$, and age group $a$ from the $Y_{w,r, a}$ true infections. Apparently, integrity for $Y_{w,r, a}$ is not guaranteed with (\ref{eq:dark}), which we could, however, impose by rounding. We further assume that $R_{w,r, a}$ and $Y_{w,r, a}$ are independent of each other, conditional on the previous week's data. We further assume  that $R_{w,r, a}$ are independent random draws for the different districts, thus the case-detection ratio may vary between the districts. Assuming further an  i.i.d.\ setting such that $\mathbb{E}(R_{w,r,a}) = \pi_{w,a}$ yields for model (\ref{eq:mean}) under \eqref{eq:dark}: 
\begin{align} \label{eq:modsimp}
\mathbb{E}\left(\tilde{Y}_{w,r, a} \vert \tilde{Y}_{w-1,r}\right) 
 &= \mathbb{E}_{R_{w}, R_{w-1}} \left( \mathbb{E}_{Y_{w}} \left( R_{w,r,a} Y_{w, r , a} | \tilde{Y}_{w-1,r}, R_{w,r,a} , R_{w-1, r}\right) \right) \nonumber \\ \nonumber
&= \mathbb{E}_{R_{w}, R_{w-1}} \Big( R_{w,r,a} \: \mathbb{E}_{Y_t}\left(Y_{w,r,a} | Y_{w-1,r}\right) \Big) \\
&= \pi_{w,a} \: \mathbb{E}_{R_{w-1}} \Big( \exp \{ \eta_{w-1, r, a}\} \Big) \: \exp\{ o_{r, a} \}  
\end{align}
where for  clarity we include the random variable as an index in the notation of the expectation. Note that 
\begin{align}\label{eq:modsimp_2}
\mathbb{E}_{R_{w-1}} \Big( \exp\{\eta_{w-1, r, a}\} \Big) 
&= \mathbb{E}_{R_{w-1}}  \left( \exp \left\{ \sum^K_{k=1} \log(R_{w-1, r , a_k}^{-1} \tilde{Y}_{w-1,r, a_k}) \theta_{a,k} + \theta_{w} \right\} \right) \nonumber \\\nonumber 
& = \exp\left\{ \tilde{\eta}_{w-1,r, a} \right\}  \mathbb{E}_{R_{w-1}}  \left( \exp \left\{ \sum^K_{k=1} \log(R_{w-1, r, k}^{-1}) \theta_{a,k}+ \theta_{w} \ \right\} \right) \\
&= \exp\left\{ \tilde{\eta}_{w-1, r, a} \ + \tilde{\theta}_{w}\right\},
\end{align}
where 
\begin{equation*}
\tilde{\eta}_{w-1,r, a }= \sum^K_{k=1} \log(\tilde{Y}_{w-1, r, k}) \theta_{a,k}
\end{equation*}
and
\begin{equation*}
\tilde{\theta}_{w} = \theta_{w} + \log \left( \mathbb{E}_{R_{w-1}} \left( \exp \left\{\sum^K_{k=1} \log(R_{w-1,r, k})\theta_{a,k}\right\} \right) \right).
\end{equation*}
Hence, combining \eqref{eq:modsimp} and \eqref{eq:modsimp_2} shows that if we fit the model \eqref{eq:lp_ts} to the observed data, which are affected by unreported cases, we obtain the same autoregressive coefficients $\theta_{a,k}$ for $k= 1,...,  A$ as for the model trained with the true (unknown) infection numbers. All effects related to undetected cases accumulate in the intercept, which is of no particular interest in this context. In summary, if we assume that the CDR does not depend on the number of infections but might be different between age groups and different weeks, we obtain valid estimates for the autoregressive coefficients even if (multiplicative) under-reporting is present. While the independence assumptions made are generally questionable, it is reasonable to assume these for a short time interval. Note that a similar argument holds for an additive CDR under epidemiological models proposed by \citet{Meyer2017} and \citet{Held2005}. 

\subsection{Infection dynamics for school children}
\label{sec:infection_dynamics_application}

\begin{figure} [t]
    \centering
    \includegraphics[width=\linewidth, page =1]{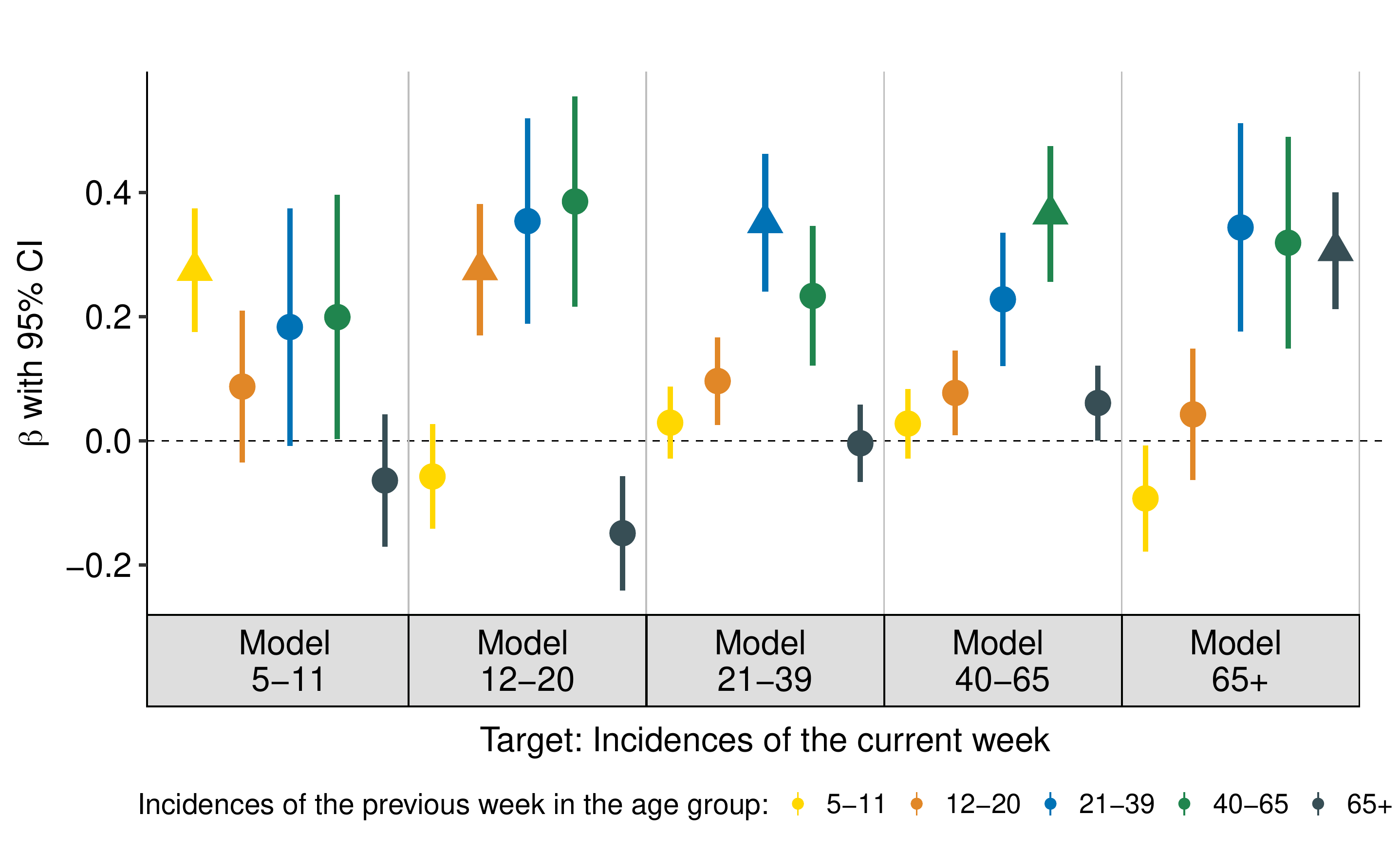}
    \caption{Association of previous week's incidences in different age groups (colour-coded) with the current-week incidences for calendar weeks 9-12 in 2020 stratified by age group.}
    \label{Fig:effects}
\end{figure}

We can now investigate the infection dynamics between different age groups to answer the question brought up at the beginning of Section \ref{sec:Autoregressive}. Since the age groups provided by the RKI are too coarse for this purpose, we rely on the data provided by the LGL  for Bavaria. For this dataset, we have the age for each recorded case, which, in turn, enables us to define customised age groups. To be specific, we define the age groups of the younger population in line with the proposal of the \citet{who_2020}: 0-4, 5-11, 12-20, 21-39, 40-65, +65. For this analysis, we estimate model  \eqref{eq:mean} with data on infections  which were registered between March 1 and March 31, 2021. The data was downloaded in May 2021; hence reporting delays should have no relevant impact on the analysis. We employ model \eqref{eq:mean} separately for all five analysed age groups to assess how all age groups affect each other. The fitted autoregressive coefficients $\theta_{a,k}$ are visualised in Figure \ref{Fig:effects} including their 95\% confidence intervals. The partition of the x-axis refers to index $a$, while index $k$, the influence of the other age groups, is indicated by the different colours and drawn from left (5-11) to right (65+). For instance, the label  ``Model 5-11'' shows all interpretable effects where the target variable is the incidence of people aged between 5 and 11. 
Note that the only interpretative results of our model concern the dependencies between the age groups. Thus we omit the weekly intercept estimates from \eqref{eq:lp_ts} in Figure \ref{Fig:effects}, which lose all interpretative power in the context of under-reporting as argued in Section \ref{sec:robustness}. 

In general, we observe that the autoregressive effects for the own age group, i.e. $a = k$ (drawn as triangles in Figure \ref{Fig:effects}) are among the essential predictors in all age-group-specific models. Regarding the effects between age groups, the association of 5-11-year-olds (yellow, most left coefficient) with all other age groups is relatively small and, in most cases, not significant. In contrast, the age groups of working people aged between 21-39 (blue, middle) and 40-65 years (green, second right) have the highest relative effect on the incidences for all age groups (except for the autoregressive coefficients). For instance,  we see that the effects of the children and adolescents (5-11 and 12-20 years) on the incidences of 21-39 and 40-65-year-olds, albeit sometimes being significantly different from 0, affect the prediction far less than the incidences of the working population. In this respect, the results confirm previous analyses concluding that increasing incidences in children and adolescents are weakly associated with the incidences of other age groups. Vice versa, we find empirical evidence that people between 21 and 65 are the main drivers of infection dynamics. 

The results do not come without limitations. First of all, note that the data is observational, not experimental. Hence, we can only draw associative and not causal conclusions from the data without additional assumptions. Moreover, we rely on the given assumptions on the under-reporting. Still, rerunning the analyses for other weeks, shown in the Supplementary Material, yielded similar results, supporting the robustness of our approach and findings. 

\section{Modelling hospitalisations accounting for reporting delay}
\label{sec:reporting_delay_hospitalization}

A relevant number of COVID-19 infections lead to hospitalisations, and the incidence of patients hospitalised in relation to COVID-19 is of paramount importance to policymakers for several reasons. Firstly, hospitalised cases are  most likely to result in very severe illnesses and deaths, the minimisation of which is generally the primary aim of healthcare management efforts. In addition, knowing the number of hospitalised patients is crucial to adequately assess the current state of the healthcare system. Finally, while the number of detected infections depends considerably on testing strategy and capacity, the number of hospitalisations provides a more precise picture of the current situation. For these reasons, hospitalisation incidence has been deemed increasingly more relevant by scientists and decision-makers over the course of the pandemic, and finally became the central indicator for pandemic management in Germany from September 2021, complementing the incidence of reported infections.

\begin{figure} [t!]
    \centering
    \includegraphics[width=0.7\linewidth, page =1]{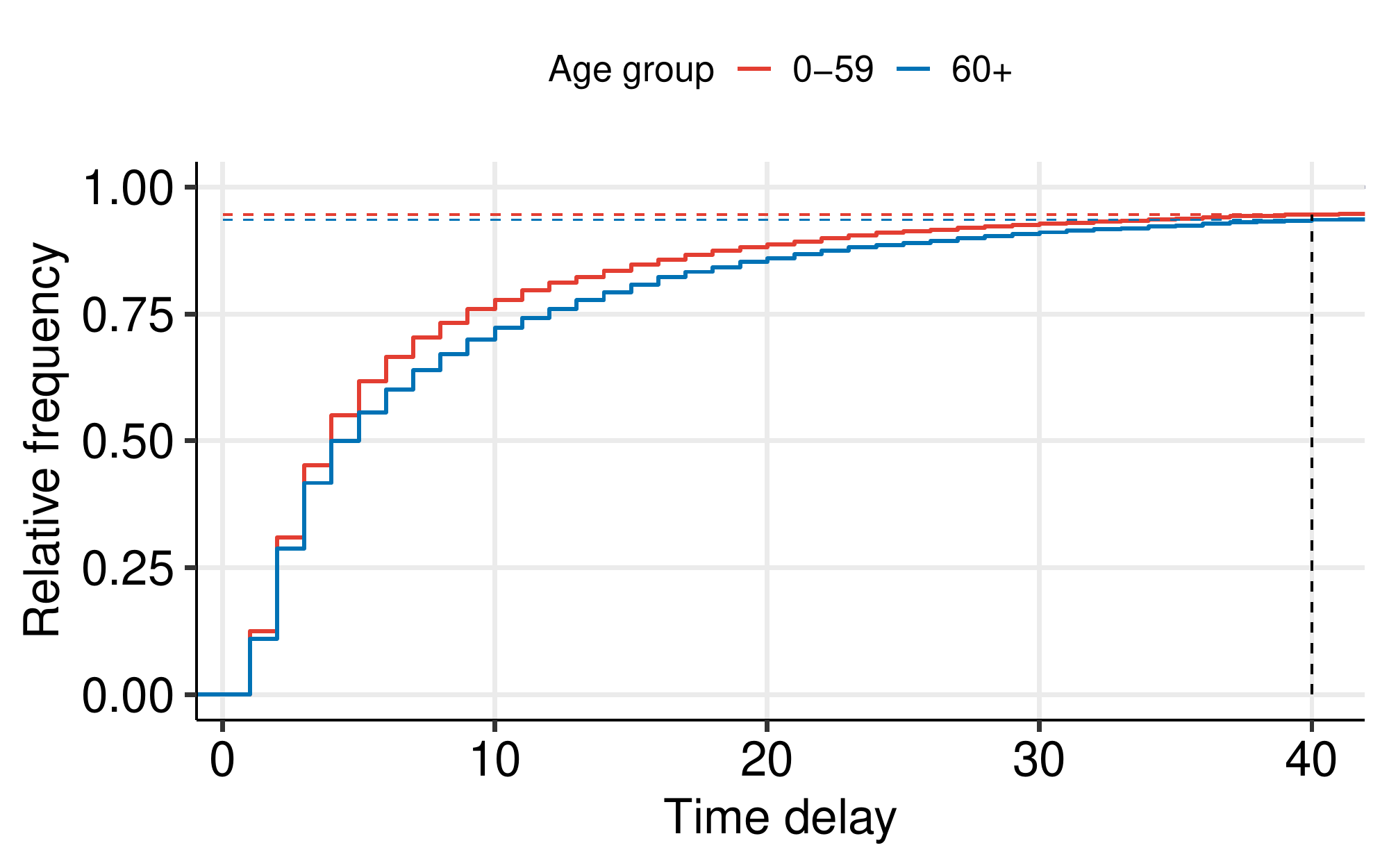}
    \caption{Cumulative distribution function of the time delay (in days) between hospitalisation and its reporting, calculated with data from January 1st to November 18th of 2021, shown separately for the age groups 0-59 and 60+. The curves for both age groups are truncated at a delay of 40 days, when approximately 94\% of all hospitalisations have been reported.} 
    
    \label{Fig:delay}
\end{figure}

The central problem in calculating the hospitalisation incidence with current data is that hospitalisations are often reported with a delay. Such late registrations occur along reporting chains (from local authorities to central registers), but also due to data validity checking at different levels. Visual proof of the degree of this phenomenon is given in Figure \ref{Fig:delay}, which depicts the empirical distribution function of the time (in days) between the date on which a patient is admitted into a Bavarian hospital and the date on which the hospitalisation is included in the central Bavarian register. In 2021, only $11.6\%$ of hospitalised cases in Bavaria are known the day after admission, and about two thirds of them ($66.4\%$) are reported within seven days. Moreover, the duration tends to be slightly shorter for patients younger than 60 than older patients. 

Modelling and interpreting current data with only partially observed hospitalisation incidences can lead to biased estimates and misleading conclusions, especially if one is interested in the temporal dynamics. To correct for such reporting delays, we utilise ``nowcasting'' techniques, loosely defined as \textsl{``[t]he problem of predicting the present, the very near future, and the very recent past''} \citep[p. 193,][]{BanBura2012}. Related methods have been extensively treated in the statistical literature (see, e.g., \citealp{Hohle2014,Lawless1994}) and successfully applied to fatalities and infections data during the current health crisis \citep{denicola:2020,Gunther2020,Schneble:2021}. In contrast to these approaches, we focus here on modelling the hospitalisation incidences, correcting for delayed reporting through a nowcasting procedure based on the work of \citet{Schneble:2021}. 

We denote by ${R_{t,r,g}}$ the hospitalisation incidence on day $t$ for district $r$ and age/gender group $g$, while the absolute count of hospitalisations in the same cohort is defined by $H_{t,r,g}$. Naturally, those two quantities related to one another through 
\begin{align}
 {R_{t,r,g}} = \frac{{H_{t,r,g}}}{x_{\text{pop},r,g}}. \label{eq:R_trg}
\end{align}
To account for the delayed registration of hospitalisations in $H_{t,r,g}$ when modelling $R_{t,r,g}$, we pursue a two-step approach consisting of a nowcasting and a modelling step. In the former step, we nowcast the hospitalisations that are expected but not yet reported, while in the latter step we model ${R_{t,r,g}}$ as a function of several covariates, which will allow us to gain insights into the geographic and sociodemographic drivers of the pandemic. We describe the two steps below.
\subsection{Nowcasting model}
\label{sec:nowcasting}
\begin{figure} [t!]
    \centering
    \includegraphics[width=0.9\linewidth, page =1]{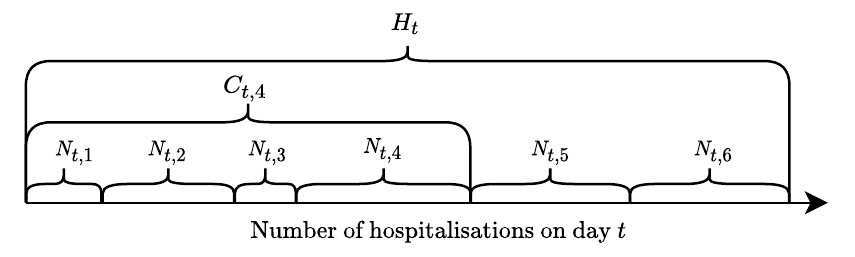}
    \caption{Illustration of the data setting for $d_{\text{max}} = 6$. $N_{t,d}$ indicates hospitalisations reported with a specific delay $d$, while $C_{t,d}$ denotes all those reported with delay {up to} $d$. $H_t$ denotes the final number of hospitalised cases regardless of the delay with which they were reported, that is with a delay up to the maximum possible, $d_{\text{max}}$.}
    \label{Fig:illustration}
\end{figure}

In this first step, we estimate the final number of hospitalised patients on day $t$, denoted by $H_{t}$,  factoring in the expected reporting delay. Note that, while we have data available at the district level, at this stage we aggregate hospitalisations across Bavaria due to the sparsity of the data. If we are performing the analysis on day $T$, we can compute the cumulative hospitalisation counts $C_{t,d} = \sum_{l = 1}^d N_{t, l}$, where $N_{t,d}$ is the number of hospitalisations on day $t$ reported with delay $d$, for every $t\in \{1, ..., T\}$ and $d \in \{1, ..., T-t\}$. Assuming a maximal reporting delay of $d_{\text{max}}$ days, we denote the complete distribution of delayed registrations of cases with hospitalisation on day $t$ by $N_{t} = (N_{t,1}, ..., N_{t,d_{\text{max}}}) \in \mathbb{N}^{d_{\text{max}}}$ with $\sum_{d = 1}^{d_{\text{max}}} N_{t,d} = H_ {t}$. We graphically demonstrate how $N_{t,d}, C_{t,d}$, and $H_{t}$ relate to one another in Figure \ref{Fig:illustration}. By design, $N_{t}$ follows a multinomial distribution: 
\begin{align}
    N_{t} \sim \text{Multinomial}(H_{t}, \pi_{t}), \label{eq:multinom}
\end{align}
where $\pi_{t} = \left(\mathbb{P}(D_{t} =1;t), ..., \mathbb{P}(D_{t} =d_{\text{max}};t)\right)$ are the proportions of hospitalisations on day $t$ with a specific delay, and $D_{t}$ is a random variable describing the reporting delay of a single hospitalisation which occurred at time $t$. For this application, we do not directly model those probabilities but instead opt for a variant of the sequential multinomial model proposed by \citet{Tutz1991}. In particular, we define the conditional probabilities through
\begin{equation}
        p_{t}(d| x_{t}) := \mathbb{P}(D_{t} = d|D_{t}\leq d; x_{t}),\label{eq:seq}
\end{equation}
conditional on covariates $x_{t}$. It follows that the cumulative distribution function of $D$ can be written as:
    \begin{align}
  F_{t}(d| x_{t}) &=  \mathbb{P}(D_{t} \le d; x_{t, a})=\mathbb{P}(D_{t}\le d_{t}| \le d+1; x_{t}) \cdot \mathbb{P}(D_{t} \le d+1; x_{t}) \nonumber  \\  &  = \prod_{k = d+1}^{d_{\text{max}}} (1- p_{t}(d | x_{t})).
    \label{eq:cumul}
\end{align}
Combining \eqref{eq:multinom} and \eqref{eq:seq} allows us to model the delay distribution with incomplete data. We do this separately for two age groups, which we denote by an additional index $a$. This leads to the model
    \begin{equation}
    N_{t,a,d} \sim \text{Binomial} \left(C_{t,d}, p_{t,a}(d | x_{t, a, d})\right)
    \label{eq:quasi_bin}
\end{equation}
with the structural assumption
\begin{align*}
\log\left(\frac{p_{t,a}(d | x_{t, a, d})}{1 - p_{t,a}(d | x_{t, a, d})}\right) &= \theta_0 + s_1(t) + s_2(d) + s_3(d)\cdot \mathbb{I} (60+) + x_{t, d}^\top\theta,
\end{align*}
where $\theta_0$ is the intercept, $s_1(t) = \theta_1 t + \sum_{l = 1}^L\alpha_l \cdot (t - 28l)_+
$ is the 
piece-wise  linear
time effect, $s_2(d)$ the smooth duration effect, $s_3(d)$ a varying smooth duration effect for the age group 60+, and $x_{t, d}$ are additional covariates depending on $t$ and the delay $d$, i.e.\ a weekday effect for $t$ and $t+d$. 



From Figure \ref{Fig:illustration}, one can also derive that the proportion of $H_{t,a}$ included in $C_{t,a,d}$ can be comprehended as the probability that a hospitalisation on day $t$ in age group $a$ has a reporting delay smaller than or equal to $d$, i.e. $F_{t,a}(d| x_{t, a})$. Assuming independence of $H_{t,a}$ from $D_{t,a}$ then yields:  
\begin{equation}
    \mathbb{E}(H_{t,a}) F_{t, a}(d| x_{t, a}) = \mathbb{E}(C_{t, a, d}),
\end{equation}
meaning that the expected number of patients from age group $a$ hospitalised on day $t$ can finally be obtained as 
\begin{equation}
    \mathbb{E}(H_{t,a}) = \frac{\mathbb{E}(C_{t, a, d})}{F_{t, a}(d| x_{t, a})}. \label{eq:nowcasting}
\end{equation}
In summary, we can fit the logistic regression model given by \eqref{eq:quasi_bin} with the available data on hospitalisations. Based on this model, we exploit \eqref{eq:nowcasting} to obtain an estimate for the expected number of hospitalisations from age group $a$ on day $t$. Uncertainty intervals for the estimated nowcasts can then be obtained e.g. through a parametric bootstrapping approach relying on the asymptotic multivariate normal distribution of the estimated model coefficients.

\subsection{Hospitalisation model}
\label{sec:hosp_model}
In the second step, we propose a model for the expected value of $R_{t,r,g}$, the hospitalisation incidence on day $t$ in district $r$ and age/gender group $g$, conditional on covariates $x_{t,r,g}$. To be specific we set 
\begin{align}
    \mathbb{E}(R_{t,r,g}| x_{t,r,g}) &= \exp\{\theta_0 + \theta_{\text{age}}x_{\text{age},g} + \theta_{\text{gender}}x_{\text{gender},g} + \theta_{\text{gender:age}}x_{\text{age},g} x_{\text{gender},g}+ \nonumber\\
    &\hspace{1.4cm}\theta_{\text{weekday}}x_{\text{weekday},t} +
    s_1(t) + s_2( x_{\text{Lon},r},  x_{\text{Lat},r}) + u_{r}\}\nonumber\\ &= \exp\{ \eta_{t,r,g}\} , \label{eq:link}
\end{align}
where the linear predictor $\eta_{t,r,g}$ includes, in addition to the intercept $\theta_0$, effects for the age/gender groups through the main and interaction effects $ \theta_{\text{age}},  \theta_{\text{gender}}$ and $\theta_{\text{gender:age}}$. Additionally, we include dummy effects $\theta_{\text{weekday}}$ for each day of the week to account for potentially different hospitalisation rates over the course of the week. Furthermore, the hospitalisation incidences are allowed to vary over time through the smooth term $s_1(t)$. Finally to account for spatial heterogeneity, we add a smooth spatial effect of each district's average longitude and latitude $s_2(r)$ and a Gaussian random effect to capture random deviations from this smooth effect, i.e. $u_{r} \sim N(0,\tau^2)$ with $\tau^2 \in \mathbb{R}^+$.

Note that, on any given day $t > T - d_{\text{max}}$, we do not yet observe the final hospitalisation counts $H_{t,r,g}$, but only the ones already reported at this time, that is $C_{t,r,g,T-t}$, indicating the cumulative observations on day $t$ in district $r$ reported with a delay of up to $d$ days for age/gender group $g$. The age/gender group indexed by $g$ extends the coarse (binary) age categorisation $a$ used in Section \ref{sec:nowcasting}, which only differentiates between cases younger and older than 60 years. Exploiting \eqref{eq:nowcasting} and the definition \eqref{eq:R_trg} of the incidence 
leads to the final model
\begin{align}
    \mathbb{E}(R_{t,r,g}| x_{t,r,g}) = \frac{\mathbb{E}(C_{t,r,g,T-d}| x_{t,r,g})}{x_{\text{pop},r,g}F_{t, g}(T-t| x_{t, g})}.
   \label{eq:c_t}
\end{align} 
We thereby  set $C_{t,r,g,T-d} = H_{t,r,g}$ if $T-d \geq d_{\text{max}}$.
Rearranging \eqref{eq:c_t} shows that modelling the count variable $C_{t,r,g,T-d}$ with the offset term  $\log(x_{\text{pop},r,g}F_{t, g}(T-t| x_{t, g}))$  is equivalent to modelling $R_{t,r,g}$ as in \eqref{eq:link}, since 
\begin{align}
    \mathbb{E}(C_{t,r,g,T-t}| x_{t,r,g}) = \exp\left\{ \eta_{t,r,g} + \log(x_{\text{pop},r,g}F_{t, g}(T-t| x_{t, g}))\right\} = \mu_{t,r,g} \label{eq:c_trg}
\end{align}
holds. In practice we thereby replace the unknown quantities in the offset through their estimates derived in the previous section. In other words, the delayed reporting is easily accommodated through an (estimated) offset in the model using only the reported data $C_{t,r,g,T-t}$. With this "trick" We can now finalise the model, where we here  make use of   a negative binomial model  to account for possible overdispersion, that is:
\begin{align*}
   C_{t,r,g,T-t}| x_{t,r,g} \sim \text{NB}(\mu_{t,r,g}, \sigma^2),
\end{align*}
with $\mu_{t,r,g}$ parametrised as in \eqref{eq:c_trg} and \eqref{eq:link}, and the dispersion parameter $\sigma^2$ is estimated from the data. 

As an additional note, we point out that accounting for late registrations works analogously for any model within the endemic-epidemic framework originating in \citet{Held2005}. The only difference to the approach presented here is that the exact functional form of the expected value must be adequately accounted for. For instance, if $\mu_{t,r,g}$ consists of the sum of non-negative endemic and epidemic terms, one should incorporate the offset in both terms.

\subsection{Application to the fourth COVID-19 wave in Bavaria}

For the application, we focus on the first two months of the fourth wave of the pandemic in Bavaria, which began towards the end of September 2021. In particular, we consider hospitalisations between September 24th and November 18th, using data reported as of November 19th, 2021. We set $d_\text{max} = 40$ days to be the maximum possible duration between hospitalisation and its reporting in the central Bavarian register. We derive this choice from the empirical delay distribution in Figure \ref{Fig:delay}, proving that since the beginning of 2021, around $94\%$ of the hospitalisations have been reported within 40 days of their occurrence.  We have no information on the date of hospital admission for about $9.6\%$ of all hospitalisations related to COVID infections that were reported between September 24th and November 19th. For those cases, we replace the date of hospitalisation with the respective COVID-19 infection date as reported by the local health authorities. For brevity, we only present a comparison of the nowcasted and raw hospitalisation counts for the nowcasting model and the age/gender group-specific and spatial effects of the hospitalisation model. We refer to the Supplementary Material for additional results. 

\begin{figure} [t!]
    \centering
    \includegraphics[width = \linewidth]{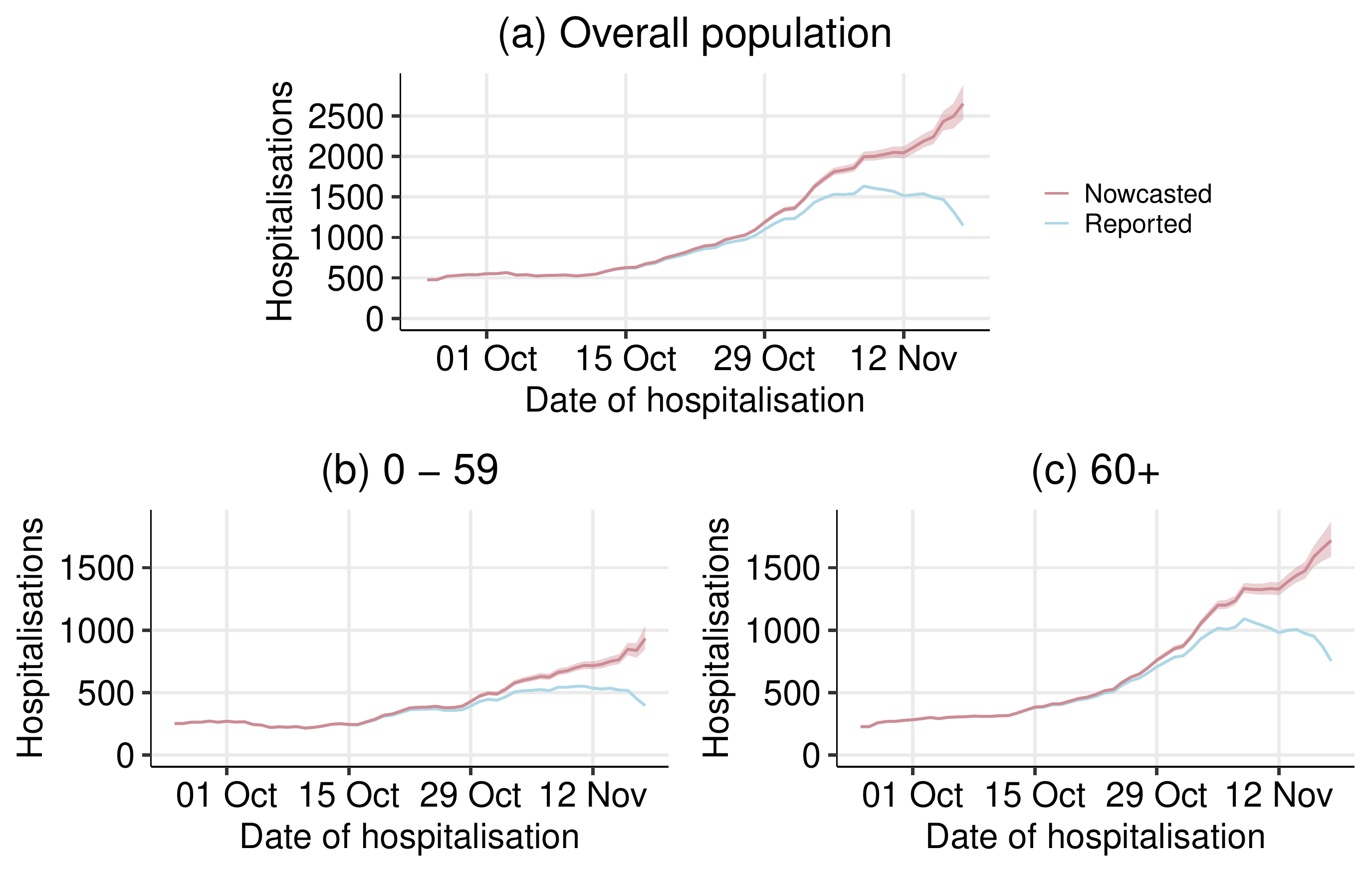}
    \caption{Comparison of nowcasted (red) and reported (blue) rolling weekly sums of hospitalisation counts between September 24th and November 18th, 2021, based on data reported as of November 19th, 2021. 95\% confidence intervals of the nowcast estimates are indicated by the shaded areas. Results are displayed for the overall population (a) as well as for age groups 0-59 (b) and 60+ (c).}
    \label{Fig:nowcast_results_sum}
\end{figure}

Figure \ref{Fig:nowcast_results_sum} maps the raw and corrected rolling weekly sums of hospitalisation counts accompanied by the $95\%$ confidence intervals for the whole population as well as separately for the two age groups under consideration. While reported numbers indicate a relatively stable or even slightly decreasing development over the last two weeks observed, the nowcast reveals a continuous upward trend since the beginning of October. Comparing both age-stratified populations, the increase for those over 60 years (the more vulnerable) is steeper. These results emphasise the need to adjust reported hospitalisation counts, as they tend to systematically underestimate the number of recently occurred hospitalisations, which can lead to inaccurate conclusions about the current situation.


Turning to the results of the hospitalisation model proposed in Section \ref{sec:hosp_model}, the estimated coefficients for all age and gender combinations can be seen in Figure \ref{Fig:spatial_hosp_effects}. Those estimates reveal considerably lower hospitalisation rates for people younger than 35 than all other age groups. We generally observe a positive correlation between age and risk of hospitalisation for both genders, i.e. older people are more likely to be hospitalised. The only exception to this intuitive finding is men over 80 years, whose expected hospitalisation rates are slightly lower than men aged 60 to 79. Statistically significant differences between men and women are visible across all age groups. While women in the youngest and oldest age group tend to have a (slightly) higher hospitalisation rate than men, the opposite holds for the other groups.

\begin{figure} [t!]
    \centering
    \includegraphics[width=0.8\linewidth, page =1]{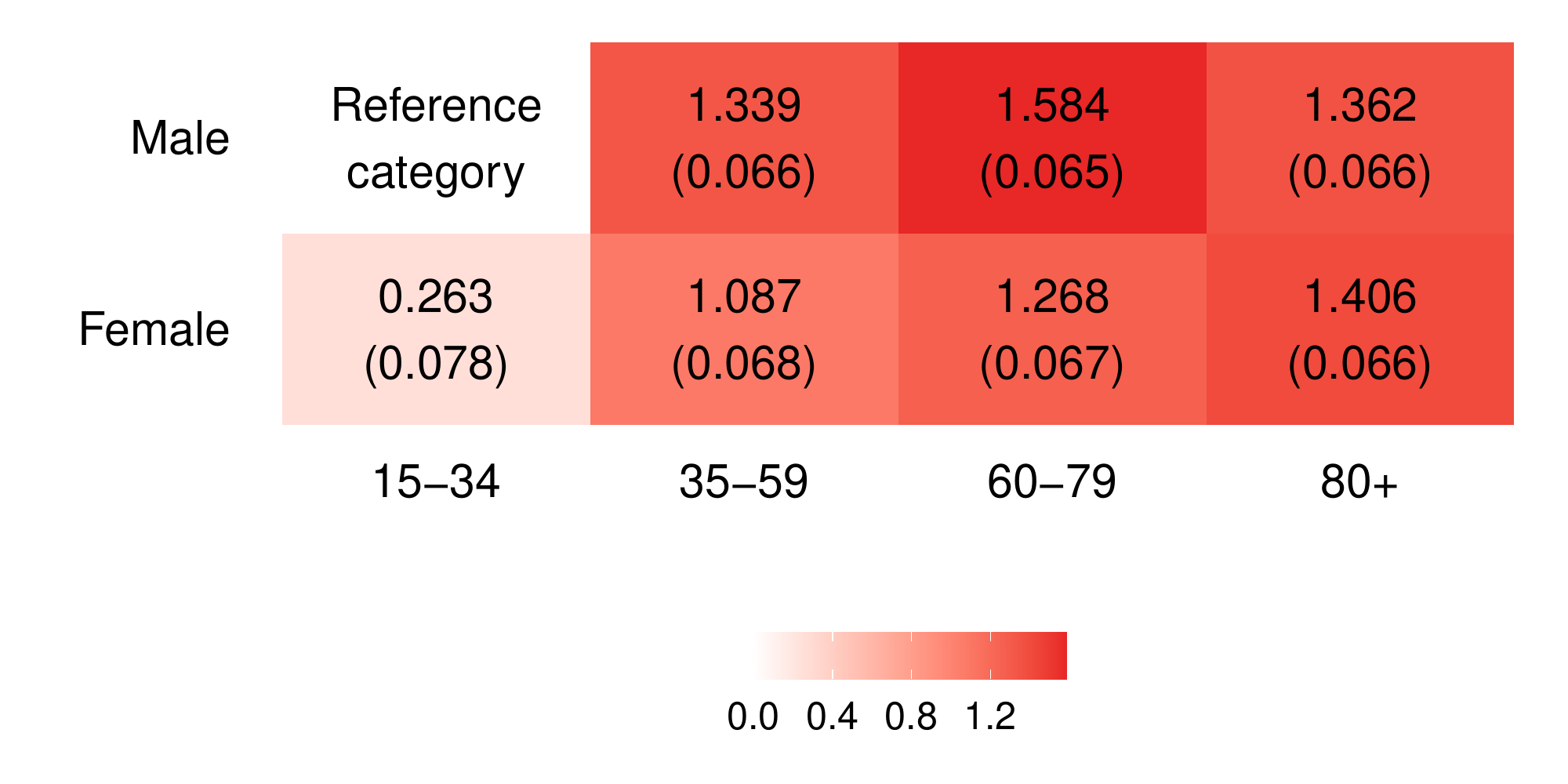}
    \caption{Estimated linear effects for different age and gender groups in the hospitalisation model, where males aged 15 to 34 are the reference category. 
    Estimated standard deviations are written in brackets.}
    \label{Fig:spatial_hosp_effects}
\end{figure}
Figure \ref{Fig:spatial_hosp_effects} depicts the random and smooth spatial effects (on the log-scale). The smooth effect in Figure \ref{Fig:spatial_hosp_effects} (a) pictures a clear spatial pattern, with generally higher hospitalisation rates in the eastern parts of Bavaria and lower rates in the north-western districts. This structure reflects the pandemic situation in Bavaria in autumn 2021, where we observed the most severe dynamics in the respective districts. Districts with unexpectedly high or low hospitalisation rates (when compared to their neighbouring areas) can be located on the map of the district-specific random intercepts in Figure \ref{Fig:spatial_hosp_effects} (b). Contrary to its role as a hotspot during the second wave in autumn 2020, the district with the lowest random effect is Berchtesgadener Land. 

\begin{figure} [t!]
    \centering
    \includegraphics[width=\linewidth, page =1]{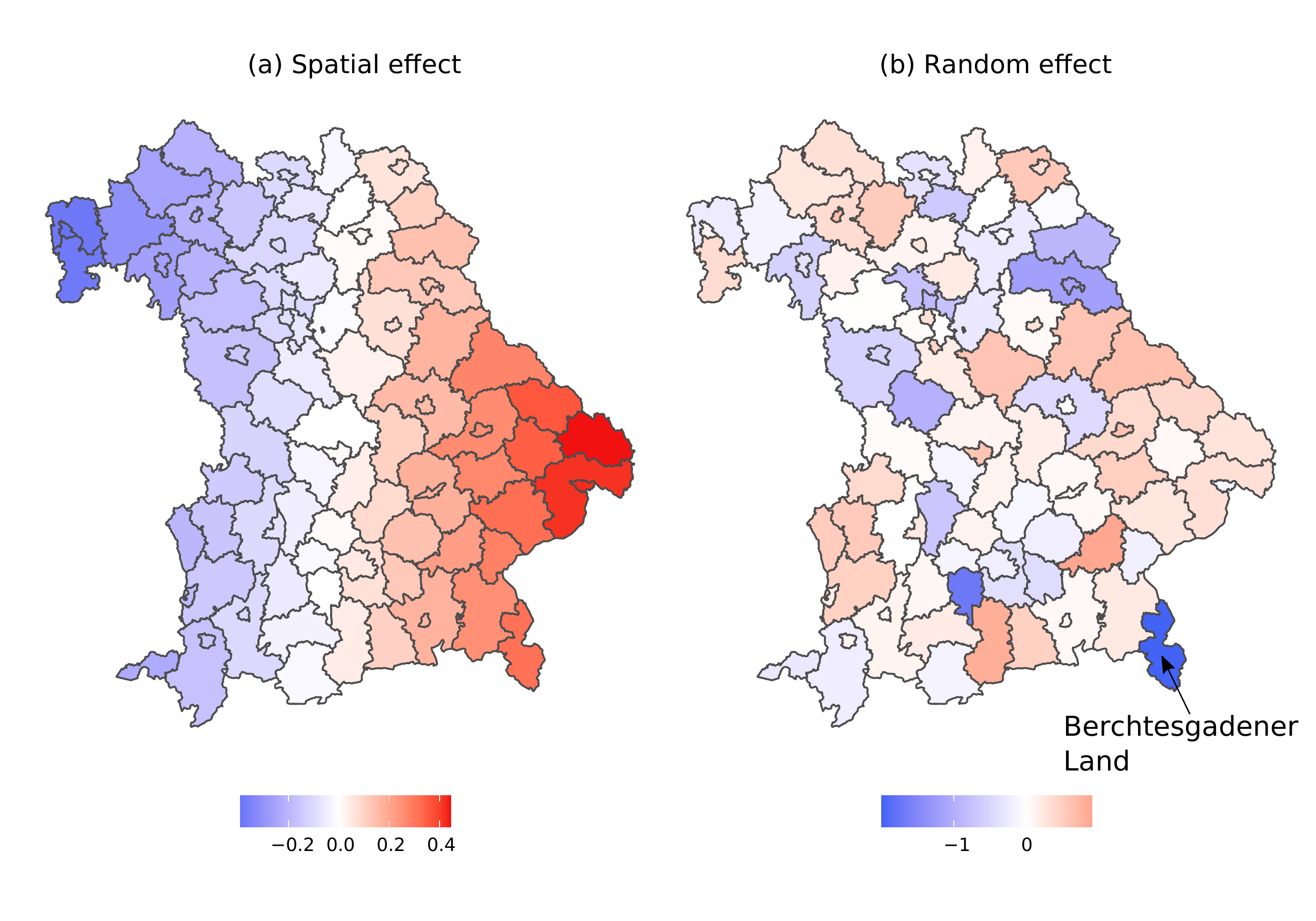}
    \caption{Estimated smooth spatial effect (a) and district-specific random effect (b) in the hospitalisation model.}
    \label{Fig:spatial_hosp_effects}
\end{figure}

\section{Modelling ICU occupancy}
\label{sec:modeling_icu}
Understanding the infections of COVID-19 is essential when aiming to attain insights into the disease's spread. However, the analysis of infections alone can only tell part of the story. In particular, focusing only on infections disregards information on the severity of the disease, which is crucial in understanding and anticipating potential strains on the health care system. This information, among other data,  can be captured by the ICU occupancy, which is the focus of our third application case. 

\subsection{Multinomial model}\label{sec:Mul_Hosp}
We consider the occupancy of ICUs where, as described in Section \ref{sec:data}, beds are categorised into the number of vacant beds ($Z_{w,r, 1}$), number of beds occupied by patients infected with COVID-19 ($Z_{w,r, 2}$), and number of beds occupied by patients not infected with COVID-19 ($Z_{w,r, 3}$). Further, we denote by $ Z_{w,r} = (Z_{w,r, 1}, Z_{w,r, 2}, Z_{w,r, 3})$ the vector of length three expressing the average number of ICU-bed occupancy in week $w$  and district $r$. The canonical GAM for this type of data is a multinomial model; hence the distributional assumption is: 
\begin{align}
    \label{eq:multinomial_z}
    Z_{w,r} \sim \mbox{Multinomial}\big(N_{w,r} , \pi_{w,r} \big), 
\end{align}
where $N_{w,r} = \sum_{j = 1}^3 Z_{w,r,j}$ is the known number of available beds in district $r$ and week $w$ and $\pi_{w,r} = \left(\pi_{w,r,1},\pi_{w,r,2}, \pi_{w,r,3}\right)$ defines the proportion of occupied beds in all respective categories. 

One advantage of this multinomial approach is that we implicitly account for \textsl{displacement effects} commonly observed for ICU occupancy data. Over time, as the number of beds occupied by patients infected with COVID-19 rise, both free beds and beds occupied by patients not infected with COVID-19 decrease almost simultaneously. In particular, the ``\textsl{displacement}'' may be caused by practices such as rescheduling non-urgent operations or other treatments which would have required an ICU stay, which were already common during the first wave of COVID-19 \citep{Stos2020}. These effects lead to negative correlations between the entries in $Z_{w,r}$, which is naturally accounted for in  model \eqref{eq:multinomial_z} as the correlation between arbitrary counts $Z_{w,r,k}$ and $Z_{w,r,l}$ is $-N_{w,d} \pi_{w,r,k} \pi_{w,r,l} ~ \forall ~ k,l \in \{1,2,3\}, k \neq l$. 

Taking the number of COVID-19 as the reference category, we effectively parametrise pairwise comparisons via 
\begin{equation} \label{eq:modmul}
\log\left(\frac{\pi_{w,r,j}}{\pi_{w,r,3}}\right)= \eta_{w,r,j}~ \forall ~ j= 1,2,
\end{equation} 
where the linear predictors $\eta_{w,r,j}$ are functions of covariates labeled as $x_{w,r}$ and defined by: 
\begin{align}
\eta_{w,r,j} =& \theta_{0,j} + \theta_{AR(1),j}^\top (\tilde{Z}_{w-1,r,1}, \tilde{Z}_{w-1,r,2})^\top + \theta_{I,j}^\top \log(Y_{w-1,r} +\delta) + \nonumber \\&s_j(x_{\text{Lon},r},x_{\text{Lat},r}) + u_{r,j} ~ \forall ~ j= 1,2,\label{eq:linpred2}
\end{align}
where $\theta_{0,j}$ is the intercept term. Further, we incorporate an autoregressive component in \eqref{eq:linpred2} by including the relative ICU occupancy observed in the previous week as a regressor. We denote the distribution of the different occupancies of the previous week as $\tilde{Z}_{w-1,d} = (Z_{w-1,r,1}, Z_{w-1,r,2}) / (\sum^3_{j=1} Z_{w-1,r,j})$, the respective effect is denoted by $\theta_{AR(1),j}$ for the $j$th linear predictor. We also let \eqref{eq:linpred2} depend on the previous week’s district- and age-specific infections per 100.000 inhabitants (incidences) denoted by $Y_{w-1,r,a}$, that are weighted by the coefficient $\theta_{I,j}~ \forall ~  j= 1,2$. 
To correct for district heterogeneity, we include Gaussian random effects, i.e. $u_{r,j}\sim N(0,\tau^2) ~ \forall~ r \in \{1, ..., R\}~ \forall ~ j= 1,2$. For smooth spatial deviations from these random effects, we add a bivariate function $s_j(\cdot, \cdot)~ \forall j= 1,2$ parametrised by thin-plate splines that take the longitude and latitude of each district as arguments (see \citealp{wood2003}, for more details). For notational brevity, let $\theta$ denote the joint parameter vector of \eqref{eq:linpred2} $\forall ~ j= 1,2$.

\subsection{Quantification of uncertainty}
As stated, the multinomial model has the beneficial property of automatically accounting for displacement effects. Note, however, that patients' expected length of stay in intensive care may exceed our time unit of one week, as the average stay of COVID-19 patents is about 13 days (see \citealp{Vekaria-etal:2021}). This means, that not all beds are completely redistributed at every time point of observation. However, apart from including the previous week's occupancy in the covariates, our proposed model does not adequately account for this stochastic variability.


We therefore pursue a Bayesian view and let $N_{w,r}$ be the number of ICU beds in district $r$ in week $w$. This number is known, and we assume that each week only a fixed but unknown proportion $\alpha$ of beds in the three categories become disposable, where $0 < \alpha < 1$. That is to say that  $\alpha N_{w,r}$ beds are redistributed among the three categories, where integrity is assumed but not explicitly included in the notation for simplicity. We assume that this new allocation is independent of the previous status of the beds and denote the newly allocated beds with the three-dimensional vector  $A_{w,r} = (A_{w,r,1}, A_{w,r,2}, A_{w,r,3})$. This setting translates to: 
\begin{equation*}
Z_{w,r} = (1- \alpha) Z_{w-1,r} + A_{w,r}.
\end{equation*}
For the newly allocated beds we still assume a multinomial model:
\begin{equation}
\label{eq:multinomial}
    A_{w,r} \sim \mbox{Multinomial}\big(  \alpha  N_{w,r} , \pi_{w,r} \big),
\end{equation}
with $\pi_{w,r}$ specified in \eqref{eq:linpred2}. Note, however, that we do not know $\alpha$ and that no information is provided in the data concerning the length of stay or the number of beds changing their status. To account for that data deficiency,  we impose a Dirichlet distribution on the vector $\pi_{w,r}$, where the prior information is determined by the available beds, i.e. 
\begin{equation}
\label{eq:Bayesmodel}
f_{\pi}(\pi_{w,r}) \propto \prod_{j=1}^3   \pi_{w,r,j} ^{(1-\alpha) Z_{w-1,r,j}}.
\end{equation}
Combining the prior \eqref{eq:Bayesmodel} with the likelihood from \eqref{eq:multinomial}, leads to the posterior
\begin{equation}
f_{\pi}(\pi_{w,r} |A_{w,d})  \propto   
 \prod_{j=1}^3    
\pi_{w,r,j}^{A_{w,r,j} + ( 1-\alpha) Z_{w-1,r,j}} = \prod_{j=1}^3   \pi_{w,r,j}^{Z_w,r,j}
\label{eq:likelihood_bayes}
\end{equation}
This, in turn, equals the likelihood resulting from the multinomial model and justifies the use of model \eqref{eq:modmul} even though not all beds are allocated weekly. Nevertheless, the central assumption of independent observations in standard uncertainty quantification in GAMs \citep{Wood2006} is violated. To correct for this bias, we substitute the canonical covariance of the estimators with the robust sandwich estimator based on M-estimators defined by: 
\begin{align}
    \mathbf{V}(\theta) = \mathbf{A}(\theta)^{-1}\mathbf{B}(\theta)\mathbf{A}(\theta)^{-1},
\end{align}
where we set $\mathbf{A}(\theta) = \mathbb{E}\left(- \frac{\partial}{\partial\theta\partial^\top \theta} \ell(\theta)\right)$, $\mathbf{B}(\theta) = \text{Var}\left(\frac{\partial}{\partial \theta} \ell(\theta)\right)$, and $\ell(\theta)$ is the logarithmic likelihood resulting from \eqref{eq:multinomial_z} or equivalently the logarithm of the posterior of \eqref{eq:linpred2}. See also \citet{Stefanski2002} and \citet{Zeileis2006}.  

%

\subsection{Application to the third wave}

We now employ the multinomial logistic regression  \eqref{eq:multinomial_z} to ICU data recorded during the third wave between March and June 2021. For the incidence data used in the covariates, we employ the RKI data; hence we set $A = 4$ and the age groups are: 15-34, 35-59, 60-79 and 80+. Further, we normalise all non-binary covariates:
\begin{align} \label{eq:norm}
{\tilde{x}}_i=\frac{x_i-\bar{x}}{\sqrt{\frac{1}{n}\sum_j^n(x_j-\bar{x})^2}}, ~~ \text{with}~~\bar{x}=\frac{\sum_j^n{x_j}}{n} 
\end{align}
 In this way, we facilitate the interpretation of associations and guarantee a meaningful comparison between the covariates. Due to space restrictions, we here only present the linear effects from \eqref{eq:linpred2} and refer to the Supplementary Material for the random and smooth estimates.

\begin{figure} [t]
    \centering
    \includegraphics[width=\linewidth, page =2]{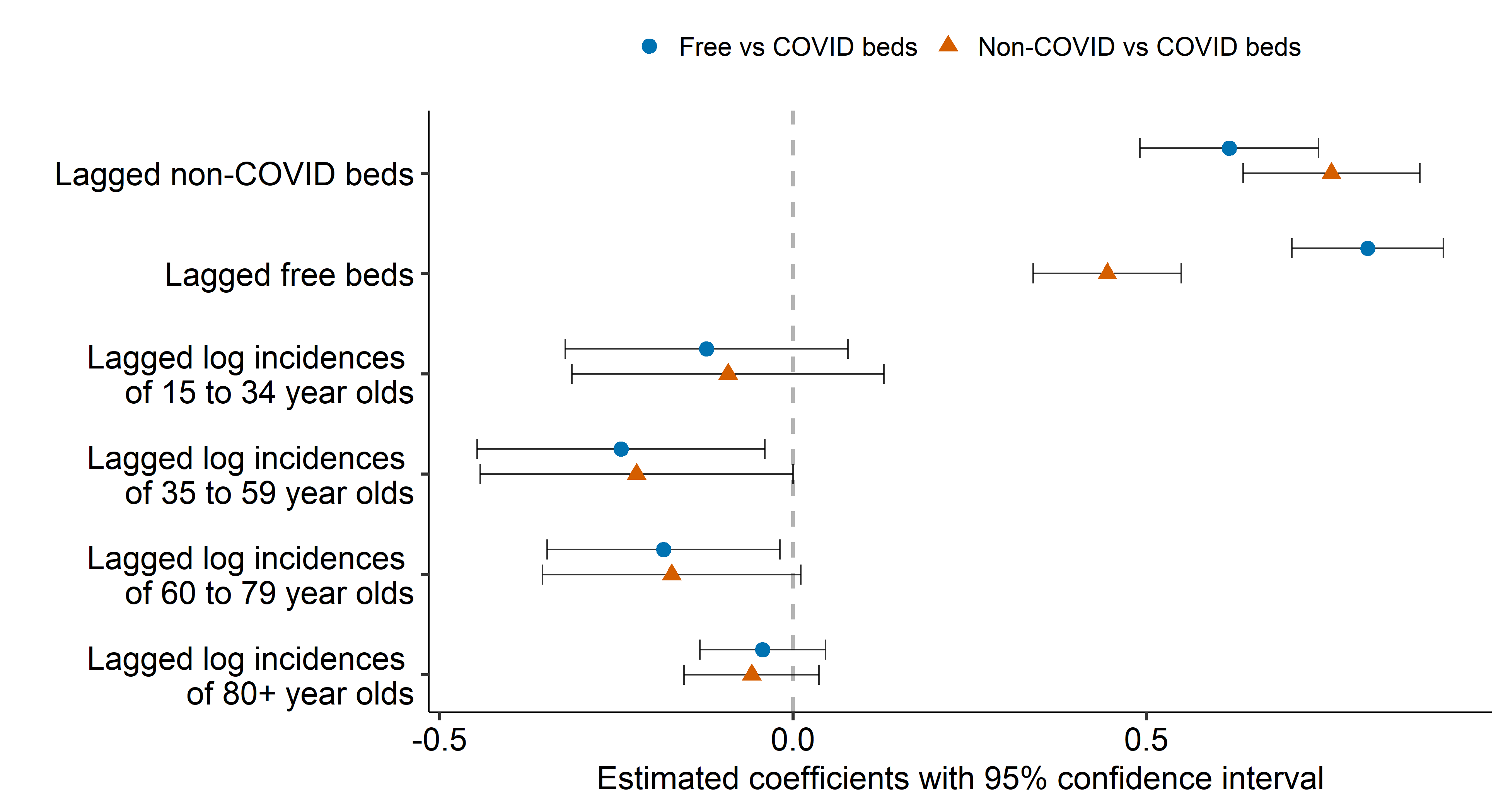}
    \caption{Estimated coefficients with confidence interval of the associations between normalised linear covariates included in the multinomial model and the logarithmic odds of a bed being free vs occupied by a patient infected with COVID-19  (blue dots) and the logarithmic odds of a bed being occupied by a patient not infected with COVID-19 vs a patient infected with COVID-19 (orange triangles).}
    \label{Fig:multi_effects}
\end{figure}

In Figure \ref{Fig:multi_effects}, we visualise the estimated coefficients including their confidence intervals. The reference category in both pairwise comparisons is COVID-beds; hence we refer to the two models by free vs. COVID beds and non-COVID vs COVID bed. In particular, the coefficients relate to the association between the covariates and the logarithmic odds of a bed not being occupied in comparison to being occupied by a patient with COVID-19, shown with blue dots in Figure \ref{Fig:multi_effects}. Analogously, the orange triangles in Figure \ref{Fig:multi_effects} illustrate the estimated association between the covariates and the logarithmic odds of a bed being occupied by a patient not infected with COVID-19 in comparison to a bed being occupied by a patient infected with COVID-19. To demonstrate the uncertainty of each estimate, a 95\% confidence interval is added. 
 Keeping the other variables constant, the normalised lagged log-incidences of all age groups generally have a negative effect on the logarithmic odds of both pairwise comparisons. This translates to the finding, that an increase in the incidences leads to a decrease in the proportion of non-COVID and free-beds in when compared to COVID beds. The lagged normalised proportion of free and non-COVID beds is estimated to have a stronger, positive association with the logarithmic odds of both pairwise comparisons. We, therefore, expect a higher number of non-COVID beds in the previous week to be followed by a higher number of non-COVID beds in the next week. 

The model can be extended to a forecasting model, as shown in the supplementary material. In particular, we demonstrate how forecasting performance changes over the different waves of the pandemic. In principle, we could also incorporate further covariates like district-specific proportions of vaccinated people. Unfortunately, these numbers are not very reliable and require sophisticated cleaning, so we prefer not to present results in this direction here.

\section{Discussion}
\label{sec:discussion}

The COVID-19 pandemic poses numerous complex challenges to scientists from different disciplines. Statisticians and epidemiologists, in particular, face the problem of extracting meaningful information from imperfect, incomplete and rapidly changing data. Generalised additive models are a powerful tool that, if used correctly, can help solving some of these challenges. In this work, we have addressed three such challenges where the utilisation of GAMs provided meaningful insight.
\begin{enumerate}
    \item We investigated whether children are the main drivers of the pandemic under a time-varying case-detection ratio.
    \item We modelled hospitalisation incidences controlling for delayed registrations, thereby providing both up to dates estimates of current hospitalisation numbers as well as insight on the demographic and spatio-temporal drivers of COVID-19.
    \item We developed an interpretable predictive tool for ICU bed occupancy that is actively used by the Bavarian government.  
\end{enumerate}
We achieved all of those results by using GAMs with different methodological extensions. Nevertheless, the use of our proposed models to extract novel information from the data provided is still subject to both data-related and methodological limitations. In general, our data sources are subject to exogenous shocks (e.g. policy changes) that lead to sudden changes in population behaviour and pose a danger to the validity of our results. Regarding the study of infection dynamics of school kids, revised testing policies hinder the long-range comparability of our findings. In the hospitalisation data, the exact date of hospitalisation is missing for about 10$\%$ of the hospitalised cases, which we impute by the given registration date of the infection. Furthermore, the records on the ICU-bed occupancy do not include intrinsic constraints, as the capacity of beds available to COVID-19 patients does not equate to the capacity of beds available to patients not infected with COVID-19. 
There are also methodological limitations. First of all, note that the data is observational, not experimental. Additionally, the set of covariates in our model can easily be extended to control for other factors, such as meteorological and socioeconomic ones. 

We close this work by emphasising that the nowcasting model can also be used as a stand-alone model. In the German COVID-19 Nowcast Hub (\cite{kit2021}), the described model is used among other nowcasting methods, including the work of \citet{Gunther2020} and \citet{kassteele2019}, to estimate hospitalisation counts on the national and federal state level in Germany. Apart from a systematic evaluation of the different approaches, one of the main goals of this project is to combine individual nowcasts to an ensemble nowcast, which may lead to more accurate estimates.
\section*{Acknowledgements}
We would like to thank Manfred Wildner and Katharina Katz on behalf of the staff of the IfSG Reporting Office of the Bavarian State Office for Health and Food Safety (LGL) for cooperatively providing the data used for sections \ref{sec:modeeling_infection_dynamics_across_age_groups} and \ref{sec:reporting_delay_hospitalization} and for fruitful discussions on the analysis of the COVID-19 pandemic. Moreover, we would like to thank all COVID-19 Data Analysis Group (CODAG) members at LMU Munich for countless beneficial conversations and Constanze Schmaling for proofreading.

\section*{Funding}
The work has been partially supported by the German Federal Ministry of Education and Research (BMBF) under Grant No. 01IS18036A. The authors of this work take full responsibility for its content. We also acknowledge support of the Deutsche Forschungsgemeinschaft (KA 1188/13-1) and the Bavarian Health and Food Safety Authority (LGL). We declare that there is no conflict of interest.

\bibliography{library}

\newpage

\section*{A Spatial units}\label{sec:SpatialUnits}

\begin{figure} [h!]
    \centering
    \includegraphics[width = \linewidth]{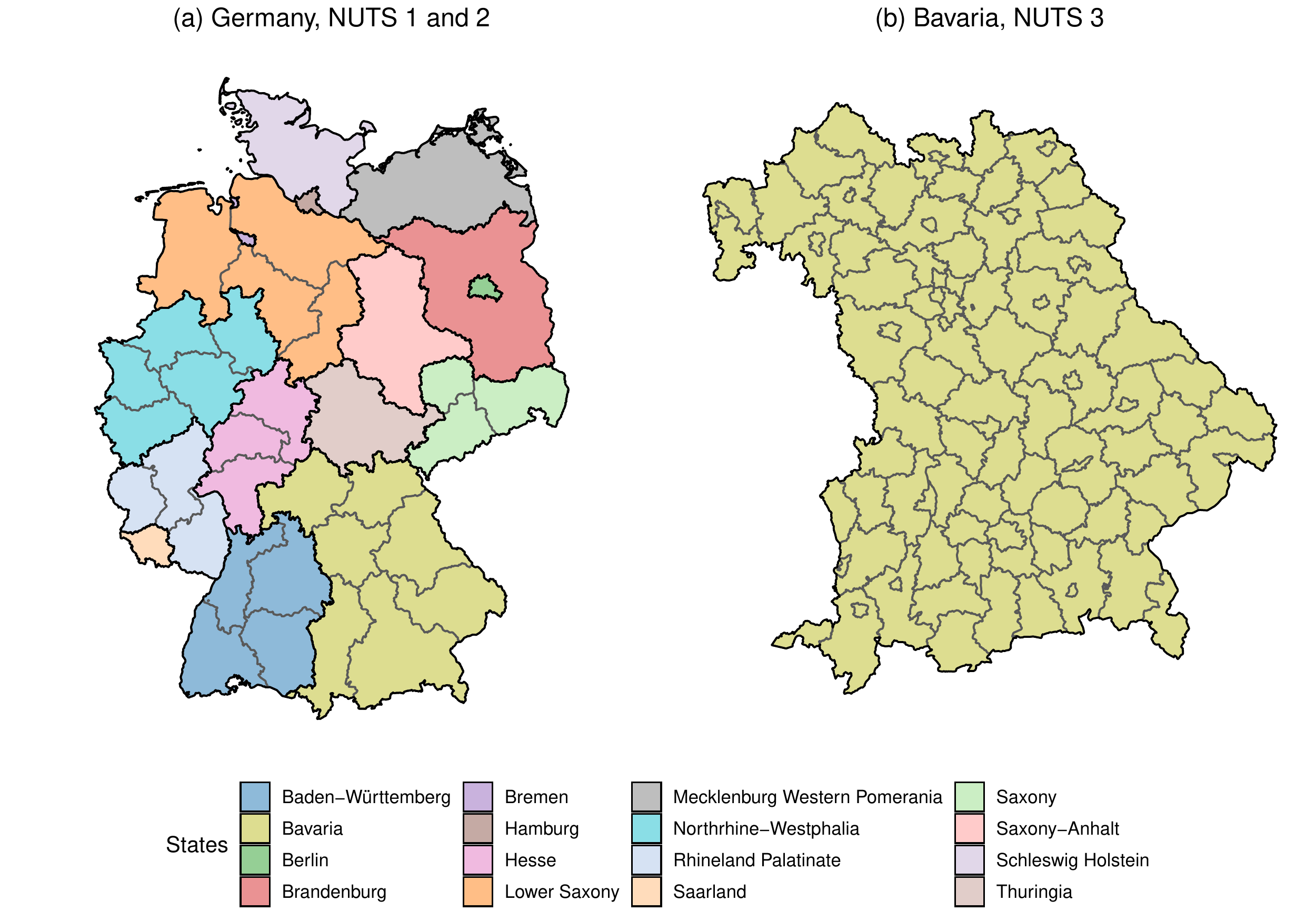}
    \caption{(a): Map of Germany, where the NUTS 1 regions are indicated by the black borders and the different colours. The NUTS 2 regions, on the other hand, are drawn in grey. Note that all NUTS 1 region borders are also NUTS 2 region borders. (b): Map of Bavaria where also the NUTS 3 regions are marked. In the legend, we state the names of each NUTS 1 region.}
    \label{Fig:nuts}
\end{figure}

We carried out most modelling endeavours presented in this article on the NUTS 3 level, which is shown on the right side of Figure \ref{Fig:nuts}.  
The only exception is the Nowcasting model from Section \ref{sec:nowcasting}, where we aggregate all data onto the NUTS 1 level in Bavaria. 
Moreover, NUTS 1 regions, depicted on the left side of Figure \ref{Fig:nuts}, are the federal states in Germany and Bavaria is one of them. In Section \ref{sec:modeeling_infection_dynamics_across_age_groups} and \ref{sec:reporting_delay_hospitalization}, we are only analysing data from Bavaria, while we employ data from complete Germany in Section \ref{sec:modeling_icu}.
\newpage

\begin{center}
	\textbf{Supplementary material: Additional results \\Statistical modelling of COVID-19 data: Putting Generalised Additive Models to work} \\
	Cornelius Fritz, Giacomo De Nicola, Martje Rave, Maximilian Weigert,  \\ Ursula Berger,     
	Helmut Küchenhoff, G{\"o}ran Kauermann \\
Ludwig-Maximilians-Universität München\hspace{.2cm}
\end{center}

\setcounter{section}{0}
\setcounter{page}{1}
\section{Analysing the association between infections from different age groups}

\begin{figure} [ht]
    \centering
    \includegraphics[width=\linewidth, page =1]{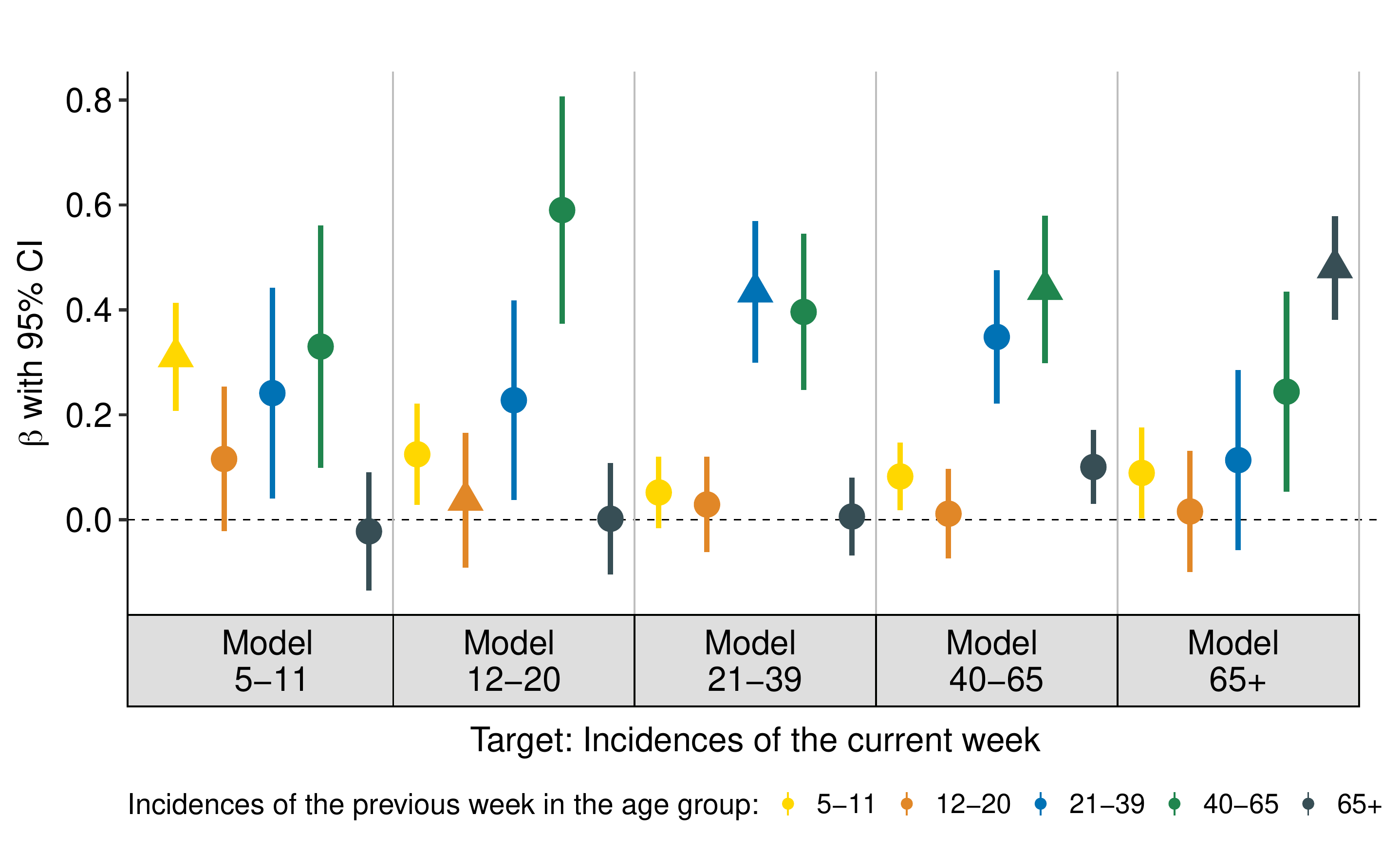}
    \caption{Association of previous week's incidences in different age groups (colour-coded marked) with the current-week incidences for calendar weeks 9-12 2020 stratified by age group.}
    \label{Fig:effects_2}
\end{figure}

To prove the robustness of the findings shown in Section 3.3 from the main article, we repeat the same analysis with data between 24.1.2021 and 29.2.2021 (calendar weeks 4 to 7). As can be seen in Figure \ref{Fig:effects_2}, the estimated coefficients are similar to the ones reported in the principal analysis of Section 3 and result in analogous interpretations. 

\newpage
\section{Modelling hospitalisations accounting for reporting delay}

\begin{figure} [ht]
    \centering
    \includegraphics[width = \linewidth]{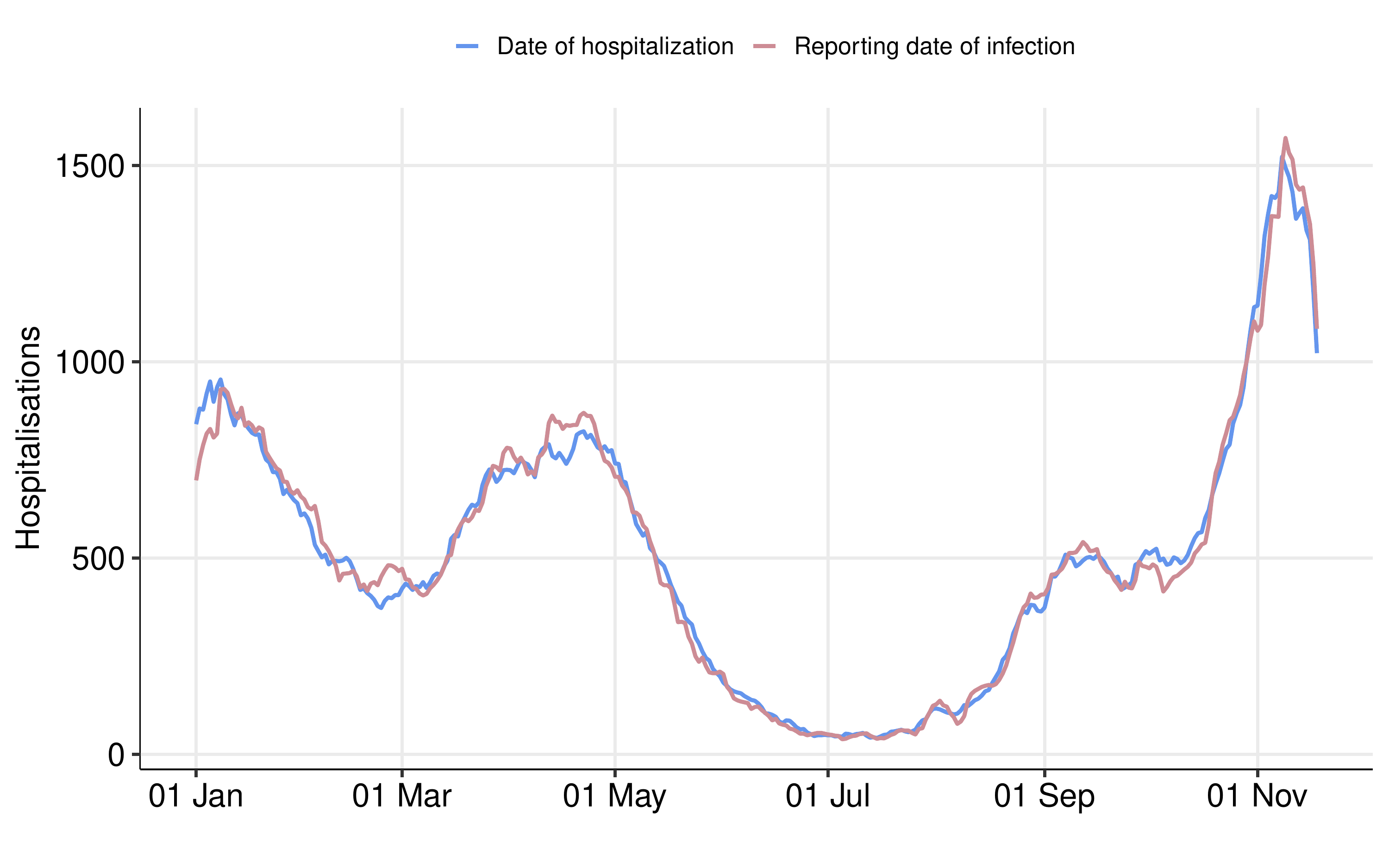}
    \caption{Comparison of the time series of the seven day sum of hospitalisations based on date of hospitalisation and reporting date of infection of hospitalised cases over the period from January 1st to November 18th, 2021. For this analysis only hospitalisations with known date of hospitalisation are included.}
    \label{Fig:comparison_meld_hosp}
\end{figure}

Modelling hospitalisations relies on the availability of hospital admission dates of COVID-19 patients. In cases where no dates are reported (about $9.6\%$ of hospitalised cases reported after September 24th), the reporting date of the infection is used instead. This choice is justified by the seven-day sum of hospitalisations over time, calculated based on both types of dates, respectively. \ref{Fig:comparison_meld_hosp} shows that there are no structural differences between both time series considering hospitalisations where both types of dates are available.

\subsection{Nowcasting model}

\begin{figure}[ht]
    \centering
    \includegraphics[width = \linewidth]{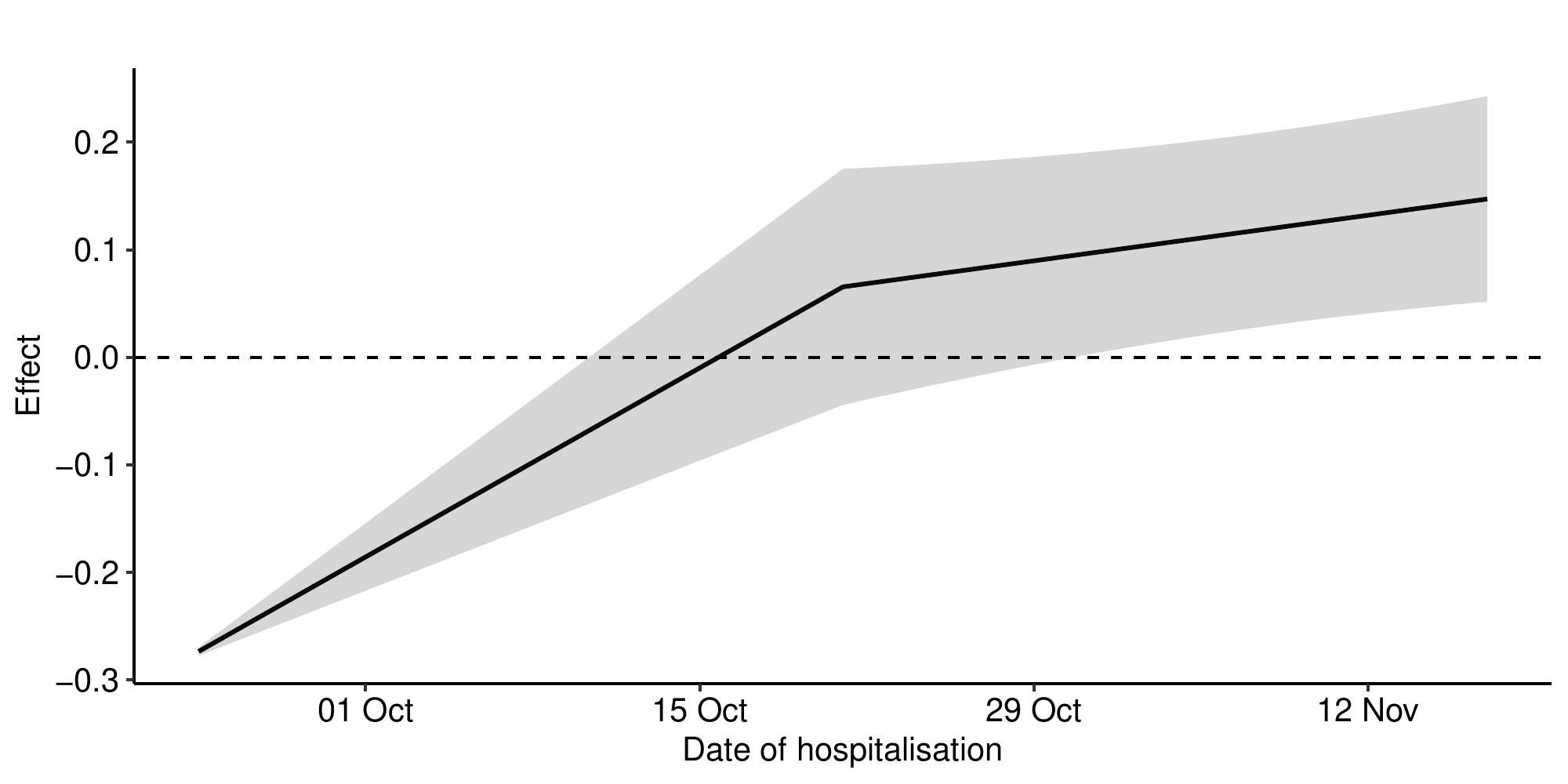}
    \caption{Estimated temporal effect within the time period between September, 24th and November 18th, 2021, accompanied by $95\%$ confidence intervals.}
    \label{Fig:temporal_effects_nowcast}
\end{figure}

\begin{figure}[ht]
    \centering
    \includegraphics[width = \linewidth]{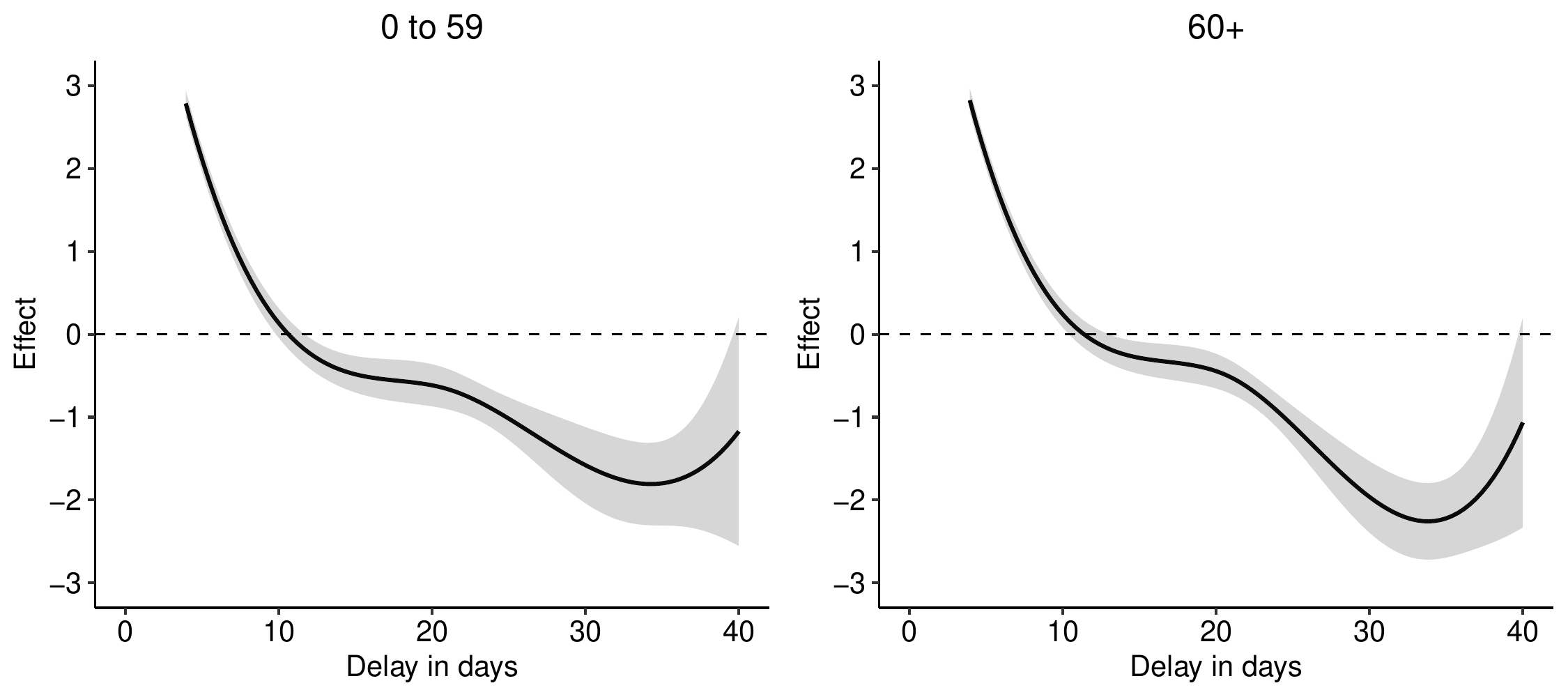}
    \caption{Estimated smooth delay effects between admission to hospital and its reporting for age groups 0-59 (left) and 60+ (right) accompanied by $95\%$ confidence bounds.}
    \label{Fig:duration_effects_nowcast}
\end{figure}

\begin{figure}[ht]
    \centering
    \includegraphics[width = \linewidth]{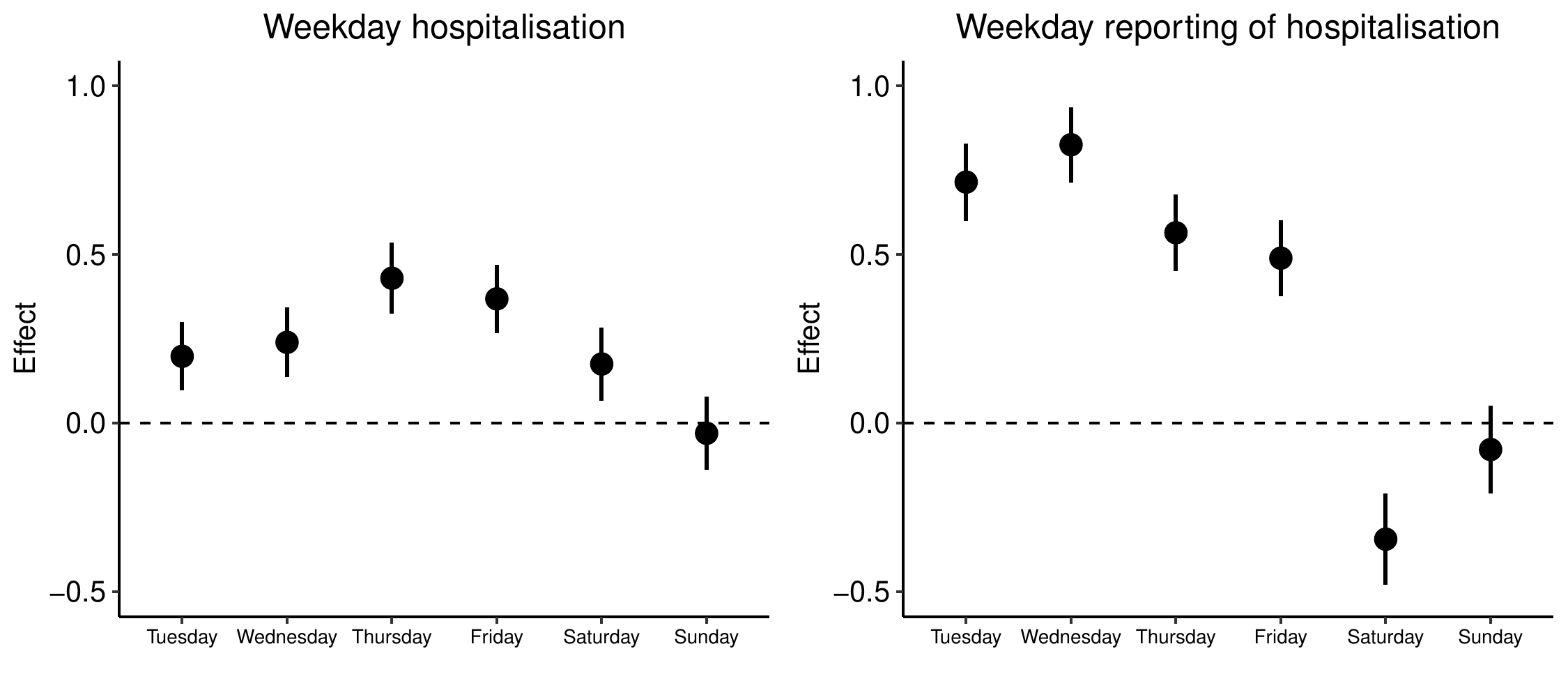}
    \caption{Estimated linear weekday effects regarding date of hospitalisation (left) and reporting date of hospitalisation (right) accompanied by $95\%$ confidence intervals. The reference category is taken to be Monday, respectively.}
    \label{Fig:weekday_effects_nowcast}
\end{figure}

In the following, we visualise all estimated effects of the nowcasting model described in Section 4.1 of the main paper. We observe the strongest association from the time delay showing a steep decrease especially in the first days after admission to hospital (Figure \ref{Fig:duration_effects_nowcast}). Comparing the effects for both age groups, only minor differences can be noticed. The piece-wise linear time effect (Figure \ref{Fig:temporal_effects_nowcast}) shows a positive trend over the considered period with a lower slope in the last four weeks.  Differences in weekdays are visible for both, the admission date to hospital as well as the date on which a hospitalised case is reported (Figure \ref{Fig:weekday_effects_nowcast}). For the latter, however, it should be noticed that new data versions of the central Bavarian register are provided from Monday to Friday only and hospitalisations newly reported on Mondays are distributed over Saturday, Sunday and Monday in a preprocessing step.

\newpage
\subsection{Hospitalisation model}

\begin{figure} [ht]
    \centering
    \includegraphics[width = \linewidth]{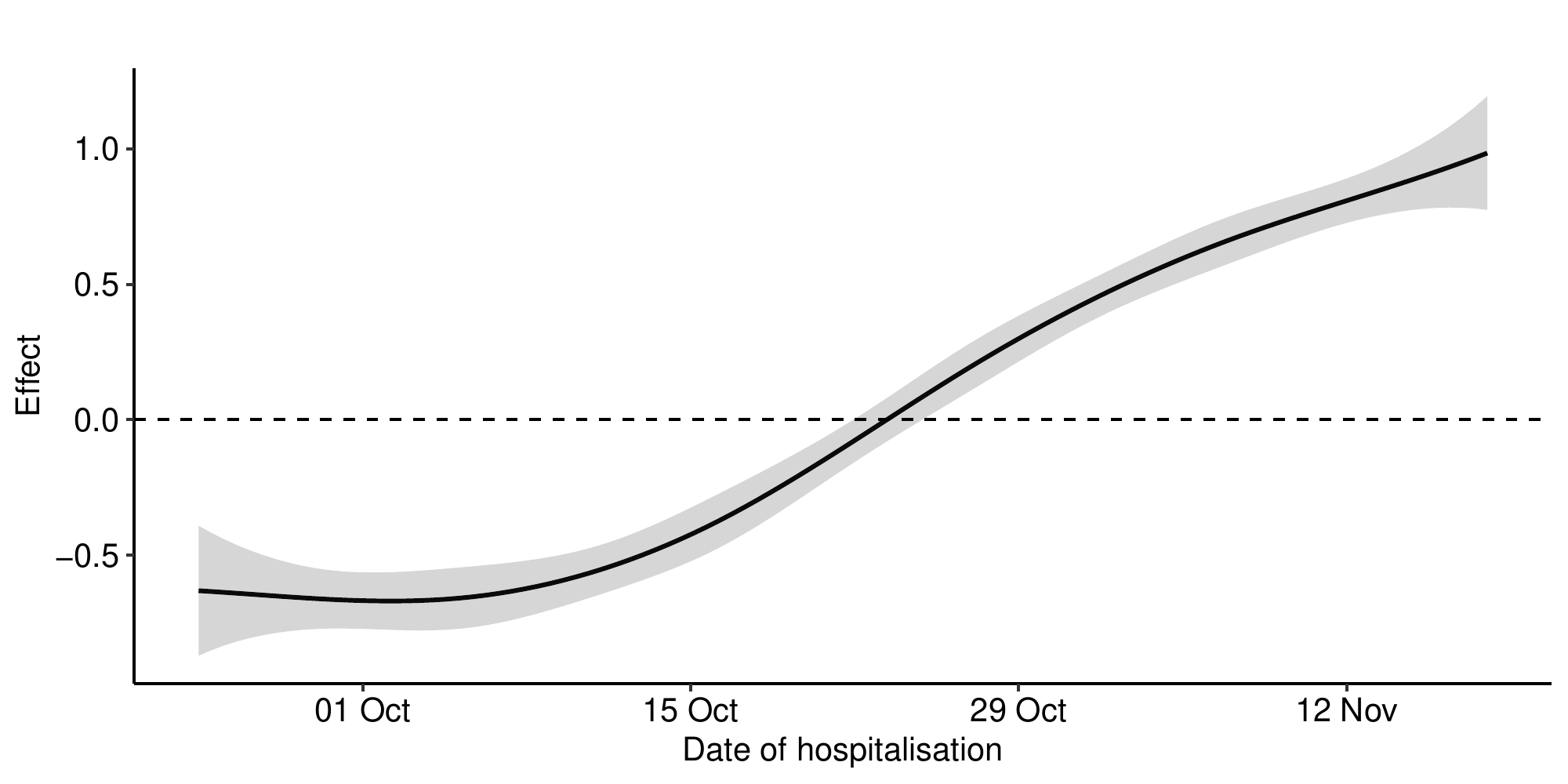}
    \caption{Estimated smooth temporal effect on hospitalisation counts within the time period between September, 24th and November 18th, 2021, accompanied by $95\%$ confidence bounds.}.
    \label{Fig:temporal_effects_hosp}
\end{figure}

\begin{figure} [ht]
    \centering
    \includegraphics[width = \linewidth]{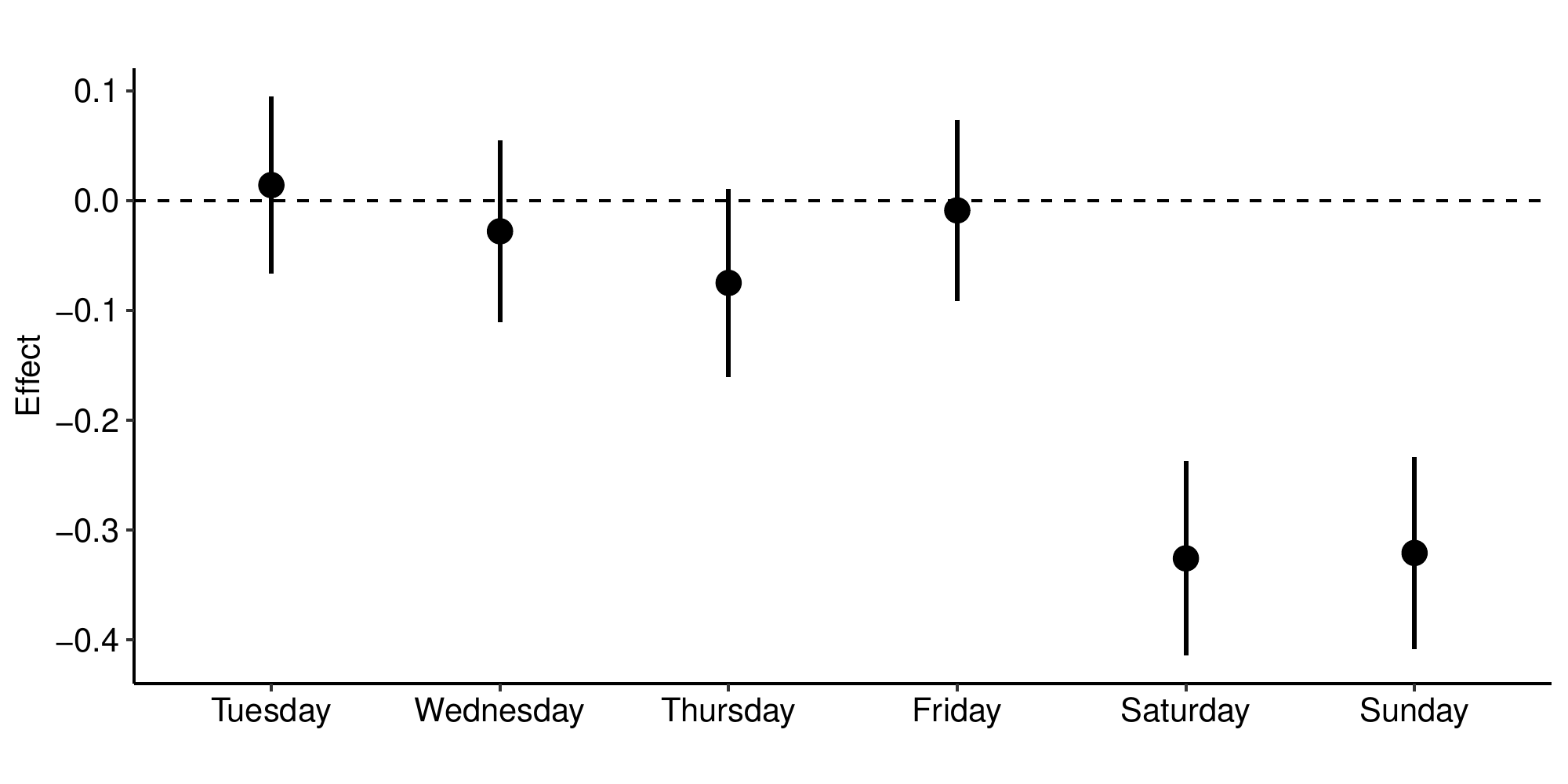}
    \caption{Estimated linear weekday effects accompanied by $95\%$ confidence intervals. The reference category is taken to be Monday.}
    \label{Fig:weekday_effects_hosp}
\end{figure}

In addition to the results described in Section 4.3, Figures \ref{Fig:temporal_effects_hosp} and \ref{Fig:weekday_effects_hosp} visualise additional effects in the hospitalisation model for which we controlled for. The smooth time indicates steady positive effects comparable to the development of the nowcasted seven day sum of hospitalisation counts illustrated in Figure 5. The weekday effects reveal a clear pattern over the week, whereas there are rather small differences from Monday to Friday and considerably less hospitalisations occur at the weekend.

\section{Modelling ICU occupancy}

\subsection{Estimates}

\begin{figure} [ht]
    \centering
    \includegraphics[width=\linewidth, page =1]{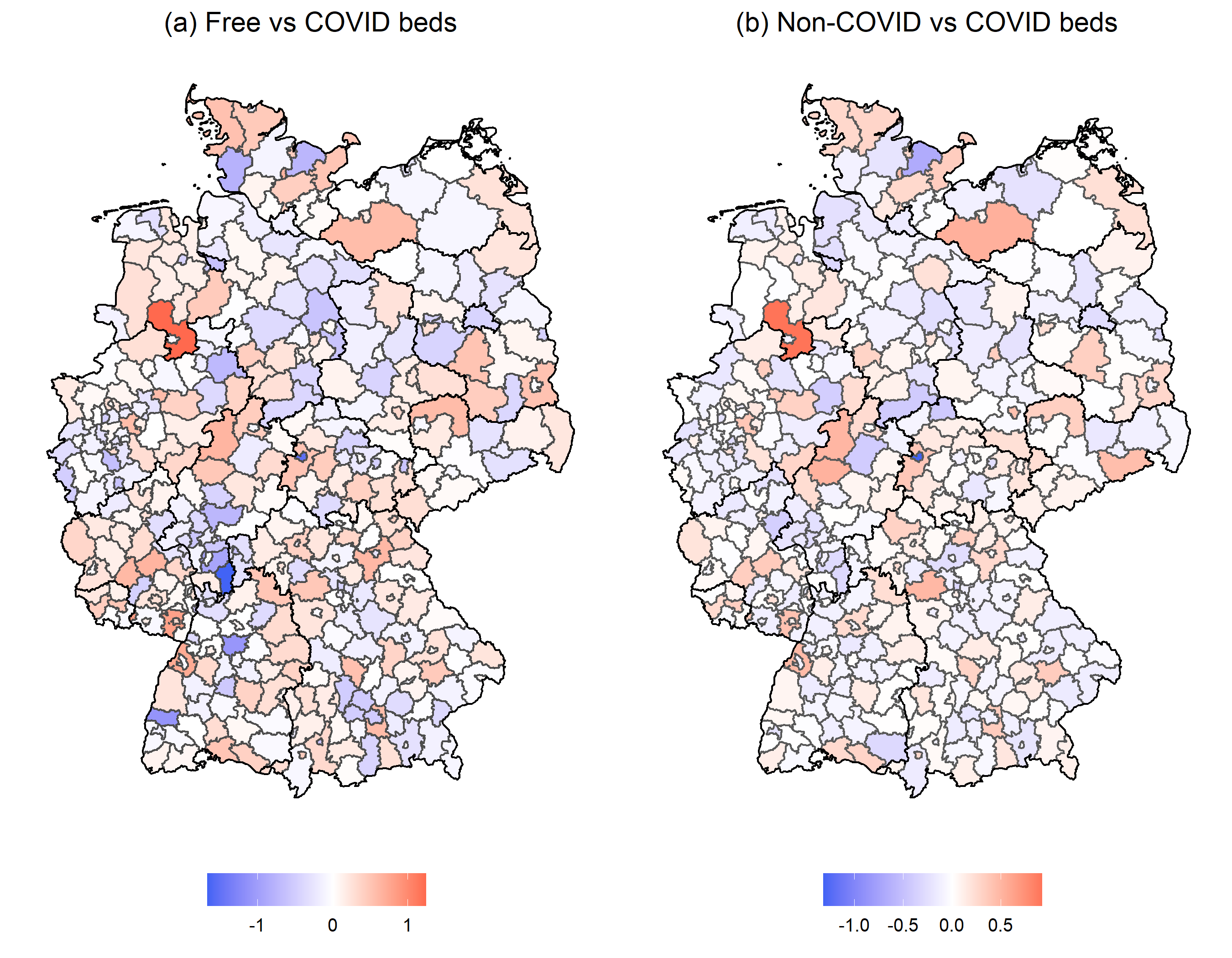}
    \caption{District specific random intercept in estimating the logarithmic odds of an ICU-bed not being occupied vs occupied by a patient infected with covid (a), and in estimating the logarithmic odds of an ICU bed being occupied by a patient not infected with COVID-19 vs a bed being occupied by a patient infected with COVID-19 (b)}
    \label{Fig:Disctrict_factor}
\end{figure}

\begin{figure} [ht]
    \centering
    \includegraphics[width=\linewidth, page =1]{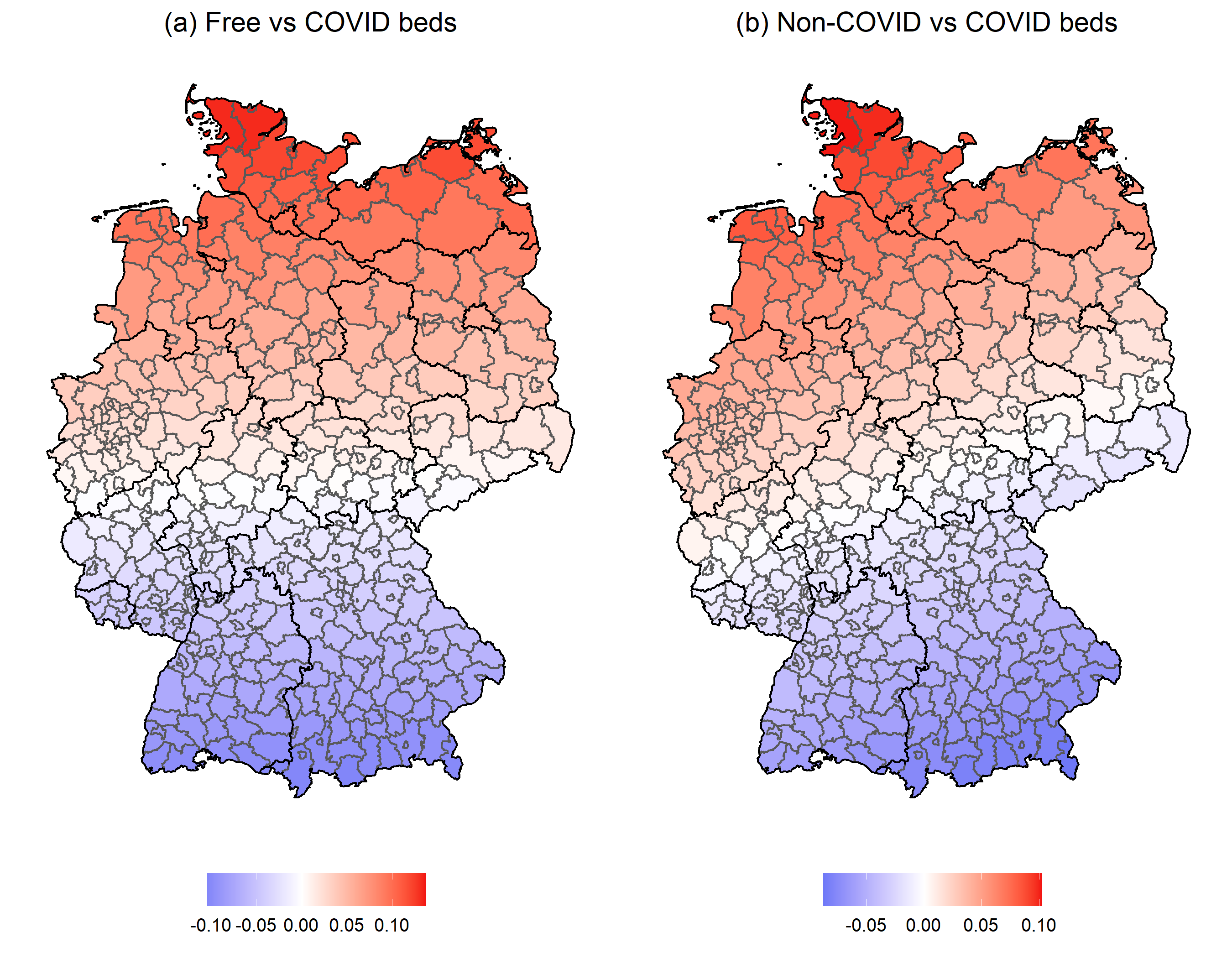}
    \caption{Estimated centred smooth spatial association with the logarithmic odds of an ICU-bed not being occupied vs occupied by a patient infected with covid (a), and with the logarithmic odds of an ICU bed being occupied by a patient not infected with COVID-19 vs a bed being occupied by a patient infected with COVID-19 (b)}
    \label{Fig:smooth}
\end{figure}

Figures \ref{Fig:Disctrict_factor} and \ref{Fig:smooth} depict the estimated district-specific random and smooth effects, respectively. Upon first inspection, there does not necessarily seem to be a clearly visible pattern in the estimated random effects, shown in Figure \ref{Fig:Disctrict_factor}. However, on a closer look, districts that include towns or even cities seem to have a lower estimated effect, while purely rural areas seem to have a higher area-specific estimated effect on the in both pairwise comparisons of the ICU occupancy.

Besides the area-specific associations estimated by our model, there could also be a spatial association in the occupancy throughout Germany.  The estimates are given in Figure \ref{Fig:smooth}. Generally, there seems to be a north-south divide, as the smoothed spatial effect appears to be higher in the north and lower in the south, where a belt from seemingly below Dresden stretching through Germany to what seems to be just below Bonn is the partition with an estimated centred effect of around zero. 

\subsection{Forecasting}

Besides interpreting the coefficients, we can use the same model to predict the occupancy of beds in the next week. To show how this works in practice, we carry out one-step-ahead predictions and evaluate the results in a rolling window framework. To obtain forecasts for the occupancy in week $w$, we train our model with the prior $8$ weeks, i.e., $w-9, ..., w-1$, and use the information on week $w$ as a test set. Since we only incorporate covariables that are lagged by one week, this setting is equivalent to providing real forecasts for the week after the end of the observational period. We let $w$ vary between the 27th of September in 2020 (calendar week 40 in 2020) and the 12th of September in 2021 (calendar week 37 in 2021). Within this time frame, we cover two critical infection waves as well as low-infection seasons. In this way, we can assess the behaviour of our predictions in many different realistic scenarios. To measure the goodness of the model, we rely on strictly proper scoring rules \citep{Gneiting2007} and use the logarithmic score as employed in \citet{Held2017} and defined by the logarithmic probability to observe the occupancy in week $w$ under the multinomial model from \eqref{eq:multinomial_z} and the predicted $\pi_{w,r}$ from the trained model. We compare the performance of the full model as specified in \eqref{eq:linpred2} with simplified specifications of the model without the lagged infection effect, without random and spatial effects, without autoregressive effects, and only including an intercept term. To further differentiate between all models, we calculate average scores for each model and employ pairwise permutation tests to compare all sub-models against the predictions under the full model as proposed in \citet{Diebolt2002}.     

\begin{figure} [t]
    \centering
    \includegraphics[width=\linewidth, page =1]{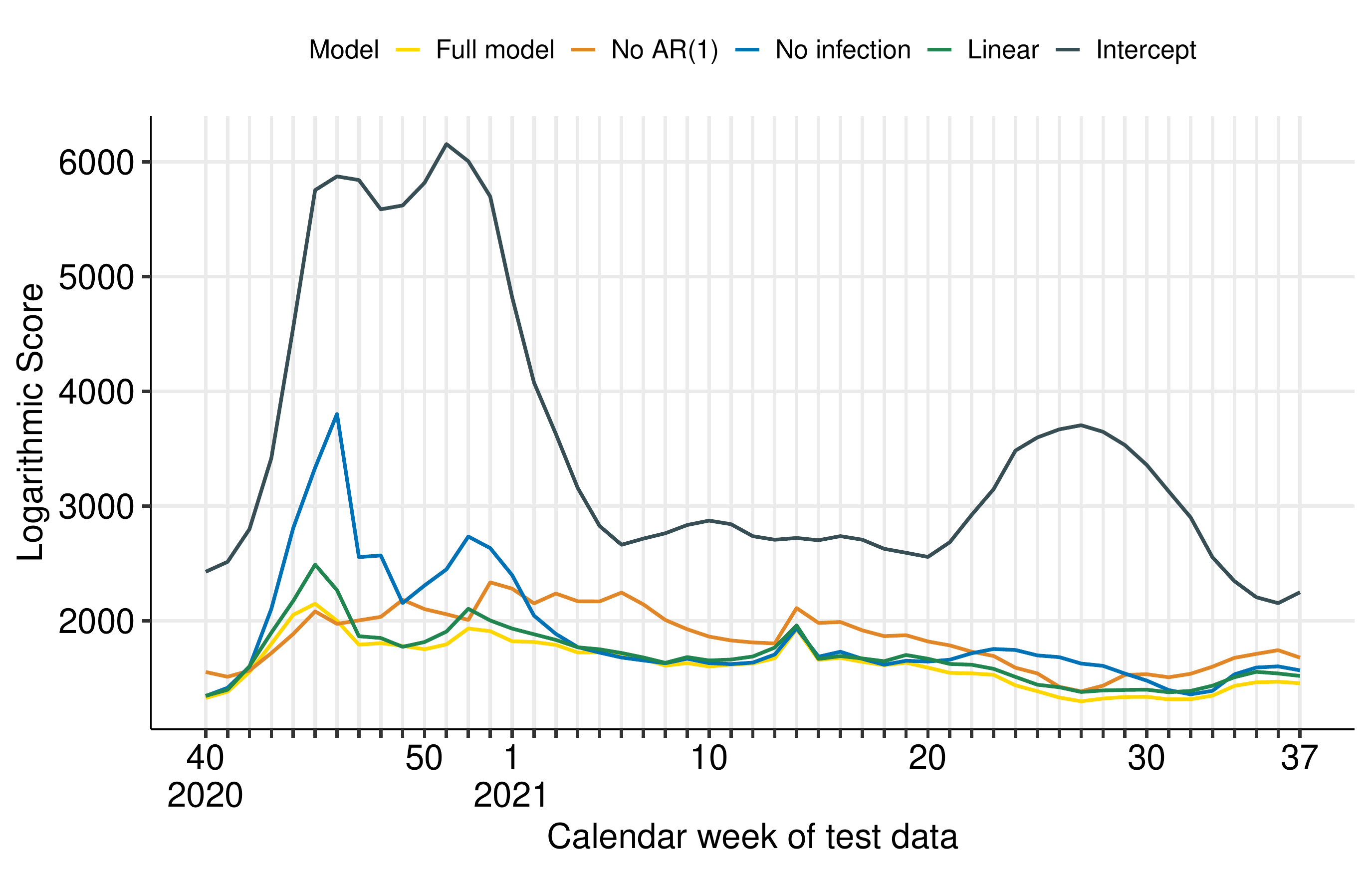}
    \caption{The logarithmic score of all candidate models from the one-week-ahead forecasts between October 2020 and September 2021.}
    \label{Fig:forecast}
\end{figure}

\begin{table}[t]
\center
  \caption{ Averaged logarithmic score and p-values from pairwise permutation tests comparing all  one-week-ahead forecasts to the full model.}
\begin{tabular}{l|ccc}
Model                         & Omitted effects & Average score & p-value         \\
\hline 
Full model    & - & 1619.243 &     \\
No AR(1)      &  $\theta_{AR(1),j}$  &1847.879 & (\textless{}0.0001) \\
No infection & $\theta_{I,j}$ & 1876.082 & (\textless{}0.0001) \\
Linear   &  $s_j(x_{r,\text{coord}}), u_{r,j}$  &1689.651 & (\textless{}0.0001) \\
Intercept   & all but $\theta_{0,j}$ & 3502.312      & (\textless{}0.0001)
\end{tabular}

  \label{tbl:prediction}
\end{table}


Figure \ref{Fig:forecast} shows the results of our one-step-ahead forecasts for all proposed models. Overall, the average score indicates that the full model provides the best performance than all other model specifications. Consistently low p-values from the permutation tests underscore this finding. We can use the intercept model to identify the high and low infection seasons regarding temporal discrepancies between the models. It is also apparent that the model performs poorly during the infection waves when leaving out the infection data, underlining that lagged infections are crucial during these periods. Not including the autoregressive component in the model, on the other hand, seems to mainly impair the logarithmic score during low-infection periods such as in calendar weeks 1 to 25 in 2021. In comparison, the random and spatial effects hardly affect the model's predictive power. However, the logarithmic score is still significantly better when including them. In summary, we can deduce from Figure \ref{Fig:forecast} that the full specification consistently provides the best forecasts for high and low infection periods.

\end{document}